\documentclass[traditabstract]{aa}

\usepackage{txfonts}
\usepackage{graphicx}
\usepackage{supertabular}
\usepackage[usenames,dvipsnames]{color}
\usepackage[colorlinks=true, linkcolor=BrickRed, citecolor=Blue, urlcolor=Blue, filecolor=Blue]{hyperref}

\newcommand{\ph}{\phantom{$-$}}

\newcommand{\oneDAV}{$\langle\mathrm{3D}\rangle$}

\newcommand{\marcs}{\textsc{marcs}}
\newcommand{\miss}{\textsc{miss}}
\def\urltilda{\kern -.15em\lower .7ex\hbox{\~{}}\kern .04em}

\newcommand{\altnai}{Na\,\textsc{i}}
\newcommand{\altnaii}{Na\,\textsc{ii}}
\newcommand{\altmgi}{Mg\,\textsc{i}}
\newcommand{\altmgii}{Mg\,\textsc{ii}}
\newcommand{\altali}{Al\,\textsc{i}}
\newcommand{\altalii}{Al\,\textsc{ii}}
\newcommand{\altsii}{Si\,\textsc{i}}
\newcommand{\altsiii}{Si\,\textsc{ii}}
\newcommand{\altphosi}{P\,\textsc{i}}
\newcommand{\altphosii}{P\,\textsc{ii}}
\newcommand{\altsuli}{S\,\textsc{i}}
\newcommand{\altsulii}{S\,\textsc{ii}}
\newcommand{\altki}{K\,\textsc{i}}
\newcommand{\altkii}{K\,\textsc{ii}}
\newcommand{\altcai}{Ca\,\textsc{i}}
\newcommand{\altcaii}{Ca\,\textsc{ii}}
\newcommand{\altfei}{Fe\,\textsc{i}}
\newcommand{\altfeii}{Fe\,\textsc{ii}}
\newcommand{\nai}{\altnai\ }
\newcommand{\naii}{\altnaii\ }
\newcommand{\mgi}{\altmgi\ }
\newcommand{\mgii}{\altmgii\ }
\newcommand{\ali}{\altali\ }
\newcommand{\alii}{\altalii\ }
\newcommand{\sii}{\altsii\ }
\newcommand{\siii}{\altsiii\ }
\newcommand{\phosi}{\altphosi\ }
\newcommand{\phosii}{\altphosii\ }
\newcommand{\suli}{\altsuli\ }
\newcommand{\sulii}{\altsulii\ }
\newcommand{\ki}{\altki\ }
\newcommand{\kii}{\altkii\ }
\newcommand{\cai}{\altcai\ }
\newcommand{\caii}{\altcaii\ }
\newcommand{\fei}{\altfei\ }
\newcommand{\feii}{\altfeii\ }

\begin{document}

\title{The elemental composition of the Sun}
\subtitle{I. The intermediate mass elements Na to Ca}

\titlerunning{Solar abundances I. The intermediate mass elements (Na to Ca)}
\authorrunning{Scott et al.}

\author{Pat Scott\inst{1}
\and
Nicolas Grevesse\inst{2,3}
\and
Martin Asplund\inst{4}
\and
A.~Jacques Sauval\inst{5}
\and
Karin Lind\inst{6}
\and
Yoichi Takeda\inst{7}
\and
Remo Collet\inst{4}
\and
Regner Trampedach\inst{8,9}
\and
Wolfgang Hayek\inst{4}}

\institute{
Department of Physics, Imperial College London, Blackett Laboratory, Prince Consort Road, London SW7 2AZ, UK\\
\email{p.scott@imperial.ac.uk}
\and
Centre Spatial de Li\`ege, Universit\'e de Li\`ege, avenue Pr\'e Aily, B-4031 Angleur-Li\`ege, Belgium\\
\email{nicolas.grevesse@ulg.ac.be}
\and
Institut d'Astrophysique et de G\'eophysique, Universit\'e de Li\`ege, All\'ee du 6 Ao\^ut, 17, B5C, B-4000 Li\`ege, Belgium
\and
Research School of Astronomy and Astrophysics, Australian National University, Cotter Rd., Weston Creek, ACT 2611, Australia\\
\email{martin.asplund@anu.edu.au, remo.collet@anu.edu.au}
\and
Observatoire Royal de Belgique, avenue Circulaire, 3, B-1180 Bruxelles, Belgium\\
\email{jacques.sauval@oma.be}
\and 
Department of Physics and Astronomy, Uppsala University, Box 516, 751 20 Uppsala, Sweden\\
\email{karin.lind@physics.uu.se}
\and
National Astronomical Observatory of Japan, 2-21-1 Osawa, Tokyo 181-8588, Japan\\
\email{takeda.yoichi@nao.ac.jp}
\and
Stellar Astrophysics Centre, Department of Physics ad Astronomy, Aarhus University, DK-8000 Aarhus C, Denmark
\and
JILA, University of Colorado and National Institute of Standards and Technology, 440 UCB, Boulder, CO 80309, USA\\
\email{trampeda@lcd.colorado.edu}
}

\date{Received 1 May 2014 / Accepted 1 Sep 2014}

\abstract{The chemical composition of the Sun is an essential piece of reference data for astronomy, cosmology, astroparticle, space and geo-physics: elemental abundances of essentially all astronomical objects are referenced to the solar composition, and basically every process involving the Sun depends on its composition.  This article, dealing with the intermediate-mass elements Na to Ca, is the first in a series describing the comprehensive re-determination of the solar composition.  In this series we severely scrutinise all ingredients of the analysis across all elements, to obtain the most accurate, homogeneous and reliable results possible.  We employ a highly realistic 3D hydrodynamic model of the solar photosphere, which has successfully passed an arsenal of observational diagnostics.  For comparison, and to quantify remaining systematic errors, we repeat the analysis using three different 1D hydrostatic model atmospheres (\marcs , \miss\ and Holweger \& M\"uller 1974) and a horizontally and temporally-averaged version of the 3D model (\oneDAV). We account for departures from LTE wherever possible. We have scoured the literature for the best possible input data, carefully assessing transition probabilities, hyperfine splitting, partition functions and other data for inclusion in the analysis.  We have put the lines we use through a very stringent quality check in terms of their observed profiles and atomic data, and discarded all that we suspect to be blended. Our final recommended 3D+NLTE abundances are: $\log\epsilon_{\mathrm{Na}} = 6.21 \pm 0.04$, $\log\epsilon_{\mathrm{Mg}} = 7.59 \pm 0.04$, $\log\epsilon_{\mathrm{Al}} = 6.43 \pm 0.04$, $\log\epsilon_{\mathrm{Si}} = 7.51 \pm 0.03$,  $\log\epsilon_{\mathrm{P}} = 5.41 \pm 0.03$, $\log \epsilon_{\mathrm{S}} = 7.13 \pm 0.03$, $\log\epsilon_{\mathrm{K}} = 5.04 \pm 0.05$ and $\log\epsilon_{\mathrm{Ca}} = 6.32 \pm 0.03$.  The uncertainties include both statistical and systematic errors. Our results are systematically smaller than most previous ones with the 1D semi-empirical Holweger \& M\"uller model, whereas the \oneDAV\ model returns abundances very similar to the full 3D calculations.  This analysis provides a complete description and a slight update of the results presented in Asplund, Grevesse, Sauval, \& Scott (2009) for Na to Ca, and includes full details of all lines and input data used.}

\maketitle
%

\section{Introduction}

The chemical composition of the Sun is a fundamental yardstick in astronomy. Essentially all analyses of the elemental abundances of planets, stars, and interstellar/intergalactic medium are referenced to the corresponding solar values.  Such analyses produce [$X$/H] values that measure differences in the abundances of species $X$ relative to the Sun.  Converting those values to the absolute abundances needed for e.g., modelling stellar and galactic chemical evolution requires knowledge of the absolute solar composition.  Having the most accurate possible estimation of the solar chemical composition is therefore of fundamental importance to all areas of astronomy, from studies of the planets of our own solar system to the most distant and ancient galaxies.  The importance of having reliable determinations of the solar elemental abundances thus cannot be overstated. Over the years, several studies have been devoted to characterising the complete solar chemical composition, from the pioneering efforts of Russell (\cite{russell29}), Suess \& Urey (\cite{suess56}) and Goldberg et al. (\cite{goldberg60}) to the more recent works of Anders \& Grevesse (\cite{ag89}), Grevesse \& Sauval (\cite{gs98}), Lodders (\cite{lodd03}), Asplund et al. (\cite{AGS05}), Lodders et al. (\cite{lodd}) and Asplund et al. (\cite{asp8}).

Unfortunately, solar abundances cannot be determined directly from the solar spectrum: the spectrum is observed, but abundances are inferred from it with modelling. This deduction involves many ingredients, all of which must be incorporated into the analysis in a realistic manner in order to achieve trustworthy results.  These ingredients include the input atomic and molecular physics, models of the solar/stellar atmospheres and spectral line formation.  For modelling late-type stars like the Sun, a particular challenge arises from the fact that the convection zone extends to the surface and thus into the spectrum-forming region. As convection is an inherently multi-dimensional and dynamic phenomenon, to model the process and its interaction with the emergent radiation field in a realistic manner requires the use of 3D hydrodynamic simulations and radiative transfer calculations (Asplund \cite{asp7}).  The first attempts at this were carried out by Dravins \& Nordlund (\cite{stell-gran4}), but computational constraints meant that it was some time before simulations reached a level of accuracy adequate for abundance analysis (see e.g. Nordlund et al. \cite{nord}).

This article is the first in a series providing the best possible determination of the solar chemical composition.  In this series we use state-of-the-art 3D atmospheric modelling, detailed calculations of line formation including departures from local thermodynamic equilibrium (non-LTE or NLTE; see e.g.\ Asplund \cite{asp7}), the most up-to-date atomic and molecular input data available, and an extremely stringent selection of photospheric absorption lines in the solar spectrum, designed to minimise blends.  In this first paper we present a comprehensive study of the solar elemental abundances of the intermediate-mass elements Na to Ca.  Accompanying articles deal with the Fe-peak elements (Scott et al. \cite{AGSS_Fegroup}; Paper II) and the heavy elements (Grevesse et al. \cite{AGSS_heavy}; Paper III). Subsequent papers will cover carbon (Asplund et al., in preparation), nitrogen (Sauval et al., in preparation), oxygen (Asplund et al., in preparation) and the light elements Li, Be and B (Asplund et al., in preparation). We presented these results in preliminary form for Li to Ca and Fe using an older 3D solar model (Asplund et al. \cite{AGS05}; AGS05), and more recently for all elements with the current 3D model (Asplund et al. \cite{asp8}; AGSS09).  The current series updates AGSS09 with further refinements of the atomic input data, line selection and NLTE line formation, and gives a full exposition of the ingredients, analysis techniques and results on which that paper was based. 

In Sect.\ \ref{s:observations} we describe the observations used in our analysis, in Sect.\ \ref{s:atmospheres} the solar model atmospheres, and in Sect.\ \ref{s:abundancecalculation} our method for calculating abundances.  We give our adopted atomic data and NLTE corrections in Sect.\ \ref{s:atomicdata}.  The results for Na to Ca are contained in Sect.\ \ref{s:results}, and are followed by discussion and conclusions in Sects.\ \ref{s:discussion} and \ref{s:conclusion}.

\section{Observations}
\label{s:observations}

We carefully analysed and measured the many lines of these elements used by Lambert \& Luck (\cite{lamb1}) and Holweger (\cite{hol1}) on three different disc-centre solar atlases.  For lines with wavelengths $\lambda$\,$<$\,1000\,nm, our adopted equivalent widths are the average of the values we measured on the Jungfraujoch (Delbouille et al.\ \cite{delb1}) and Kitt Peak (Neckel \& Labs \cite{neck}) visible solar atlases.  For $1000$\,$<$\,$\lambda$\,$<$\,1250\,nm, our equivalent widths are the average of measurements on the Jungfraujoch visible atlas and the Kitt Peak near-IR atlas (Delbouille et al.\ \cite{delb2}).  Above 1250\,nm, our equivalent widths are based exclusively on the Kitt Peak near-IR atlas. We measured our equivalent widths by directly integrating the line profiles whilst carefully accounting for any blends (i.e. in cases where small blends could be identified and accurately quantified, we did so, and did not include their contributions in our measured equivalent widths). We integrated our modelled profiles over the same spectral regions as the observed ones. 

Wherever possible, we were extremely demanding with regard to the quality of the lines retained for our analysis, carefully examining the shape and full width of each line for any trace of blending.  We gave each line a weight from 1 to 3, depending on the estimated uncertainty on our measured equivalent width, which was further modified in some cases to account for uncertainties in the atomic data (Sect. \ref{s:atomicdata}).

\section{Solar model atmospheres and spectral line formation}
\label{s:atmospheres}

Any solar or stellar abundance determination is only as accurate as its modelling ingredients. It is therefore paramount that we employ the most realistic model atmospheres possible. Here we use the same 3D hydrodynamic solar model atmosphere as in AGSS09, which was computed with a custom version of the {\sc stagger} code originally developed by Nordlund \& Galsgaard (\cite{stagger}). In this simulation, the equations for the conservation of mass, momentum and energy were solved together with the radiative transfer equation at each time-step, for a representative $6 \times 6 \times 3.8$\,Mm$^3$ volume of the quiet solar atmosphere.  This volume typically encompasses on the order of 10 convective granules at any given time.  The total temporal extent of the solar simulation sequence we use here is about 45\,min solar time, extracted in snapshots taken every 30\,s from a numerically-relaxed section of the full simulation.\footnote{Our relaxation process ensures flux constancy with height, a converged effective temperature, no drifts in the bottom boundary inflows, a nearly hydrostatic vertical stratification, and minimal p-mode amplitude (we extracted extraneous energy from the simulation by damping radial p modes during the relaxation phase). We also iteratively recomputed the opacity binning during the relaxation, based on the new $\tau$- and time-averaged atmosphere, in order to ensure convergence of the radiative transfer scheme (cf.\ Sect.~2.4 of Trampedach et al.\ \cite{trampedach:3Datmgrid}).}  This time period covers 5--10 typical granule splitting/merger timescales.  Together with the fact that $\sim$10 granules are simulated at a time, this ensures that our final temporally- and spatially-averaged line profiles are stable estimates of the integrated solar spectrum (Asplund et al.\ \cite{asp1}).

The simulation extends far below the optical surface ($\log \tau_{\rm Ross}>7$ at the bottom boundary), with the upper boundary located sufficiently far away from the region in which the lines we use are formed that numerical boundary effects do not influence our results. The top and bottom boundaries are open, and periodic boundary conditions are imposed in the horizontal directions.  The mesh is Cartesian with a numerical resolution of $240^3$, with an equidistant horizontal spacing and a vertical depth scale optimised to resolve regions of steepest temperature gradients near the surface.  Atmospheric variations on scales smaller than the grid spacing have very little impact on the resultant line profiles. Results from a precursor to the \textsc{stagger} code were shown to be essentially converged already at resolutions of $100\times100\times82$ (Asplund et al.\ \cite{Asplund2000}).  A similar convergence study with modern solar surface simulations is in progress (Collet et al., in prep.), but a recent comparison with two other state-of-the-art simulations carried out at different resolutions to our own (Beeck et al. \cite{beeck2012}) showed very good agreement between all three codes, indicating that the current resolution is indeed quite sufficient for abundance analysis.  The main reason for this is that the spectral line formation is heavily biased towards the upflows, which are divergent flows where turbulence and thus small-scale motions are less important (Asplund et al.\ \cite{Asplund2000}).

To describe the full monochromatic opacity variation as accurately as possible, during the simulation the radiative transfer was solved for 12 opacity bins sorted in both wavelength and opacity strength (Nordlund \cite{nordlund82}; Pereira et al.\ \cite{pereira_models}; Magic et al.\ \cite{magic_stagger}).  The mean stratification and rms variations of our 3D solar model can be found in Table \ref{table:3Dmodel} as a function of optical depth at a wavelength of 500\,nm ($\tau_{500\,{\rm nm}}$).  Here all averages and rms variations are calculated over surfaces of common $\tau_{500\,{\rm nm}}$, as described in Pereira et al.\ (\cite{pereira_models}).

For calculating 3D spectral line formation, we interpolated the original 3D solar simulation to a finer vertical depth scale by ignoring the optically thick lower part, similar to what is done for each time-step during the 3D hydrodynamic simulation to compute the radiative heating and cooling rates; although atmospheric variations on scales smaller than the 3D simulation have little impact on resulting line profiles, working from a finer vertical grid aides in obtaining a stable numerical solution to the radiative transfer equation.  We also downsampled the simulation horizontally for computational speed.  We have checked that the former procedure removes numerical noise without adding systematic effects, and that the horizontal downsampling has no effect on the line profiles.  We carried out line-formation calculations assuming LTE, i.e. we took Boltzmann and Saha distributions for the level populations and assumed the source function to be Planckian. A major advantage of a 3D analysis is that the traditional free parameters of solar and stellar spectroscopy (mixing length parameters for convection, and micro- and macro-turbulence for line formation) are no longer necessary (e.g. Asplund et al. \cite{asp1}). We computed the 3D LTE line formation for 45 snapshots (i.e., every second snapshot from the original simulation sequence), which is more than sufficient to obtain statistically stable line profiles when averaging; the resulting effective temperature of these snapshots is $T_{\rm eff} = 5778 \pm 5$\,K, in excellent agreement with the solar value.

For the purposes of comparison, and to estimate systematic errors in our abundance results, we also performed line-formation calculations with a series of 1D model atmospheres:\\
\begin{tabular}{p{1cm}p{7cm}}
HM: & The widely-used, semi-empirical, 1D hydrostatic photospheric model of Holweger \& M\"{u}ller (\cite{holm}). We adopted only the tabulated $T(\tau)$ stratification and obtained gas pressures, densities and electron pressures from enforcing hydrostatic equilibrium using our opacity and equation-of-state packages for the same chemical composition as the 3D model.\\
\end{tabular}
\begin{tabular}{p{1cm}p{7cm}}
\oneDAV: & A mean 3D model obtained by averaging the 3D model over surfaces of equal optical depths $\tau_{500\,{\rm nm}}$, and in time.\\  
\end{tabular}
\begin{tabular}{p{1cm}p{7cm}}
\marcs: & A theoretical 1D hydrostatic model (Gustafsson et al.\ \cite{bgus:atmgrid}; Asplund et al.\ \cite{asp0}; Gustafsson et al.\ \cite{marcs08}) extensively used for analysis of solar-type stars.\\
\end{tabular}
\begin{tabular}{p{1cm}p{7cm}}
\miss: & A semi-empirical 1D hydrostatic model (Allende Prieto et al.\ \cite{alle}) obtained from a spectral inversion of \fei and \feii line profiles. Essentially a modern-day, improved version of the HM model. It has a temperature structure very similar to what the HM structure would have looked like if the resolving power of the solar spectrum originally used by Holweger (\cite{hol0}) had been higher. 
\end{tabular}\\
For all 1D models, we adopted a microturbulence of 1 km s$^{-1}$. To ensure consistency, we performed all 1D LTE calculations using the same spectrum synthesis code as in 3D. 

Finally, we have computed 1D NLTE line formation for several elements using the {\sc multi} code (Carlsson \cite{multi}). The code relies on the same {\sc marcs} package as used for the opacity and equation-of-state calculations in both the 3D solar simulation and the 3D line formation, which minimises systematic errors when applying the resulting NLTE abundance corrections to our 3D LTE abundance results to obtain our final, recommended 3D+NLTE\footnote{By `3D+NLTE' we mean 3D LTE abundances corrected with NLTE offsets calculated in 1D, as opposed to `3D NLTE', which implies a full NLTE calculation in 3D.} results.

In many of these NLTE calculations the effect of inelastic collisions with neutral hydrogen atoms is important, but rather uncertain. Full quantum mechanical calculations have only been carried out for a few cases (e.g. Barklem et al.\ \cite{barklem03}; \cite{barklem_na}; \cite{barklem_mg}).  In the absence of such calculations, it is common to scale the classical results of Drawin (\cite{dra}) by the free parameter $S_{\rm H}$, in order to match some aspect of the available observations.  In particular, the `true' value of $S_{\rm H}$ (i.e. the one that would correctly reproduce the quantum-mechanical result) varies across different elements, lines and formation environments.  We comment on the most appropriate values of $S_{\rm H}$ for each of the elements below. 

\section{Abundance calculations}
\label{s:abundancecalculation}

We averaged simulated line intensity profiles over the temporal ($\approx 45$\,min) and spatial ($6\times6$\,Mm$^2$) extent of the 3D model atmosphere, and compared the results with the observations described in Sect.\ \ref{s:observations}.  We inferred elemental abundances from interpolation between the equivalent widths of three theoretical profiles calculated with abundances differing by \mbox{0.2 dex}.

For relevant lines, we included isotopic and hyperfine structure (HFS) in our radiative transfer calculations as a series of blending features.  For isotopic structure, we distributed the total $gf$-value of each line amongst individual components on the basis of the abundances of the respective isotopes (Rosman \& Taylor \cite{IUPAC98}), approximating the solar isotopic composition as equal to the terrestrial composition for these elements. We included HFS as described in detail in Appendix \ref{s:isohfs}.

We primarily restrict our analysis to disk-centre intensity rather than flux spectra for a number of reasons.\begin{enumerate}
\item Typically the lines are formed in deeper layers in intensity than in flux, and are therefore less sensitive to departures from LTE or the structure of higher atmospheric layers (which are more challenging to model than lower ones).
\item Line profiles are narrower in intensity, which makes the detection of possible blends easier but the achievement of agreement with the observed line profiles considerably more challenging; disk-centre intensity thus represents the most stringent test of the accuracy of solar spectral modelling.
\item Intensity profiles show effects of inhomogeneities far more clearly, including overall line shifts and C-shaped line bisectors (asymmetries), because of their narrower profiles and reduced range of formation depths compared to flux profiles.  We exploit such features to identify small hidden blends.
\item There is no need to broaden the predicted line profiles further due to solar rotation when working in disk-centre intensity.
\item Several extremely high-quality solar intensity atlases are available in both the optical and IR.
\item Computational speed; our solar abundance analysis in this series encompasses all elements and several molecules, resulting in a total of over $3000$ spectral lines.
\end{enumerate}

\subsection{Error estimations}
\label{s:errors}
Using the results from the different 3D and 1D model atmospheres, we attempted to quantify three systematic errors that might be present in our final abundance determinations: errors due to the mean temperature structure, atmospheric inhomogeneities, and departures from LTE.  We estimated the uncertainty associated with the mean temperature structure to be half the difference between the \oneDAV\ and HM results.  We took the error due to the impacts of inhomogeneities to be half the difference between the 3D and \oneDAV\ results.  The systematic uncertainty due to NLTE corrections we estimated as half of the predicted NLTE correction, bearing in mind the dependence of NLTE corrections on the poorly-known efficiency of inelastic H collisions.  Where this error term was less than 0.03\,dex, we simply took 0.03\,dex as an estimate of the minimum uncertainty, given how much even a computed NLTE correction of zero might be modified by the exact choice of input data.  As per AGSS09, we added these three uncertainties in quadrature to obtain the total systematic error.  For species exhibiting many usable spectral lines, to obtain the total uncertainty on the mean abundance we then added this systematic error in quadrature with the statistical error, which we take to be the standard error of the mean.  For species with only one or two good lines available, we instead added the systematic error in quadrature with the uncertainty stemming from the measured equivalent widths.

\section{Atomic data and line selection}
\label{s:atomicdata}

The individual transition probabilities, isotopic splittings and HFS data we have adopted are discussed in detail in the following subsections.  We give our chosen spectral lines along with their excitation energies, oscillator strengths, sources of oscillator strengths, measured equivalent widths and adopted weightings in Table~\ref{table:lines}.  Our adopted HFS and isotopic data are given in Table~\ref{table:hfs} along with relevant references.  Wavelengths of isotopic components come directly from relevant literature, whereas HFS components come from literature hyperfine constants $A$ and $B$ (cf.\ Appendix \ref{s:isohfs}).  We give our adopted partition functions and ionisation energies in Table~\ref{table:partition}.  The partition functions follow Barklem \& Collet (in preparation), and agree well with values computed from NIST atomic energy levels. The ionisation energies are from the NIST data tables.

We used damping constants for the collisions with neutral H atoms from Anstee \& O'Mara (\cite{anst}), Barklem et al.\ (\cite{bark1}), and Barklem (\cite{bark2}) where available. When these were not available we used the classical Uns\"old (\cite{unsold}) formula with an enhancement factor of two; in most such cases the lines are weak and insensitive to the adopted broadening parameter.

\subsection{Sodium}
Because of its very low ionisation energy, 5.14\,eV, Na is essentially \naii throughout the photosphere.  No line of \naii is available in the solar spectrum. A few \nai lines are present, but some of these are blended and/or too strong to be considered good indicators of the solar abundance. Starting from the selection by Baum\"uller et al.\ (\cite{bau1}), we avoided strong lines, selecting the five, mostly weak \nai lines listed in Table~\ref{table:lines}.

No modern damping calculations for collisions of Na with neutral H atoms are available for our adopted lines, so we have used an enhancement factor of two relative to the classical Uns\"old (\cite{unsold}) value, as mentioned above. As our lines are rather weak, this should not be a significant source of uncertainty.

No experimental $gf$-values are available for our solar lines; the theoretical results of Froese-Fischer \& Tachiev (\cite{FF}) are the best available data (see also the latest NIST review by Kelleher \& Podobedova \cite{kel1}).  Sodium consists entirely of $^{23}$Na, which has nuclear spin $I=\frac32$ and therefore exhibits HFS.  Fortunately, HFS data are available for nearly all the atomic levels involved in the lines we use; the best come from Das \& Natarajan (\cite{Das08}), Safronova et al.~(\cite{Safronova99}) and Marcassa et al.~(\cite{Marcassa98}).

NLTE calculations have been performed by Lind et al. (\cite{lind_na}), using realistic quantum mechanical calculations by Barklem et al. (\cite{barklem_na}) for inelastic collisions with hydrogen, instead of the dubious classical Drawin (\cite{dra}) formula almost always adopted in NLTE studies. Here we have extended this to include 1D NLTE computations for the HM, \marcs\ and \oneDAV\ 1D model atmospheres; we adopted the HM results for \miss\ and \oneDAV\ for the 3D case. Given the relatively small NLTE abundance corrections in 1D, we do not expect this assumption to be seriously in error (see e.g.\ Lind et al. \cite{lind_li6} for a full 3D NLTE study of metal-poor stars). 
Our new NLTE calculations agree well with the previous study of Shi et al. (\cite{shi1}), which was based on Drawin H-collisions.

\subsection{Magnesium}
Although \mgii is by far the dominant ionisation stage in the solar photosphere, a rather large number of seemingly clean and relatively weak lines of both \mgi and \mgii are available in the solar spectrum.  We took the line list of Zhao et al.\ (\cite{zhao1}) as the starting point for our own line selection (see Table~\ref{table:lines}). We treated the last two \mgi lines of our sample as multiplets; details can be found in the notes of Table~\ref{table:lines}.  Modern damping calculations are not available for our adopted \mgi or \altmgii\ lines.

The $gf$-values for \mgi are notoriously uncertain.  For all but two lines, we used theoretical values from the Opacity Project, obtained in LS-coupling (Butler et al.\ \cite{butl}); for the other two, we took theoretical $gf$-values from Chang \& Tang (\cite{Chang90}).

For \altmgii, we mostly adopted $gf$-values from the recent NIST review by Kelleher \& Podobedova (\cite{kel1}).  These are means of two results obtained by Froese-Fisher \& Tachiev (\cite{FF}) using different theoretical techniques, and a third theoretical result from Siegel et al (\cite{Siegel98}).  These data should be fairly accurate, as there is very good agreement between the three different theoretical results; one is assigned a rating of `A' and the other four `A+' in NIST, indicating uncertainties typically better than $\pm$0.01\,dex. The oscillator strength for the 1009.2095\,nm line comes from Kurucz's theoretical data (Kurucz, \cite{kur}). Although this dataset is generally rather inaccurate at the level of individual transition probabilities, particularly for weak transitions, in this case the $gf$-value should be quite reliable because the line is rather strong.

We specifically computed NLTE abundance corrections with for our chosen \mgi and \mgii lines, for the HM, \marcs\ and \oneDAV\ 1D model atmospheres.  For the 3D case we adopted the \oneDAV\ result. For neutral lines, our adopted Mg atom accounts for new quantum mechanical calculations of inelastic Mg+H collisions (Barklem et al. \cite{barklem_mg}); for \mgii we adopt the classical Drawin (\cite{dra}) formula with a scaling factor of $S_{\rm H}=0.1$.  For \mgi the NLTE corrections are negligible ($\le$$0.01$\,dex, as shown by Zhao et al. \cite{zhao1}; Abia \& Mashonkina \cite{abia}; Andrievsky et al. \cite{andrievsky_mg}), whereas they are significant for \altmgii, especially for the 921.8 and 924.4\,nm lines ($\approx$$-0.07$\,dex for disk-centre intensity).  We have no lines in common with the NLTE study of Mashonkina (\cite{mashonkina_mg}), but her calculations confirm that departures from LTE are insignificant for \mgi in the Sun.

\subsection{Aluminium}
We retained seven quite weak \ali lines (Table~\ref{table:lines}). We note that Al is essentially all \alii in the solar photosphere.  Accurate damping calculations are not available for \altali, but this is of little consequence for our results.

\ali transition probabilities have been discussed by Kelleher \& Podobedova (\cite{kel2}).  The data for our adopted lines come from theoretical calculations by the OP (Mendoza et al.\ \cite{mend}), under the assumption of LS-coupling.  

Al consists entirely of $^{27}$Al, which has nuclear spin $I=\frac52$.  Good HFS data exist for the lower levels of all \ali lines we consider, and can be found in the studies of Nakai et al.\ (\cite{Nakai07}), Otto \& Zimmermann (\cite{Otto69}) and Sur et al.\ (\cite{Sur05}).  In the case of the $3d\ ^2D_{3/2}$ level, the value $A=99$\,MHz from Otto \& Zimmermann (\cite{Otto69}) was confirmed by Zhao et al.\ (\cite{Zhao86}), but the measured value of $B$ is not large enough to be statistically distinguishable from zero; we therefore set \mbox{$B=0$} for this level.  HFS data are only available for two of the upper levels, from Belfrage et al.\ (\cite{Belfrage84}) and St\"uck \& Zimmermann (\cite{Stueck70}).

NLTE studies for the Sun in 1D have been made by Baum\"uller \& Gehren (\cite{bau2}) and by Gehren et al.\ (\cite{gehr}) for the solar flux spectrum. These works show that the NLTE corrections are very small. T. Gehren (2010, private communication) has kindly recomputed the NLTE corrections for our lines (see Table~\ref{table:lines}), albeit in flux and using a 1D theoretical MAFAGS-OS\footnote{This model is based on opacity sampling (OS), and is a revision of the earlier MAFAGS-ODF models of Fuhrmann et al.\ (\cite{Fuhrmann97}), which instead employed opacity distribution functions (ODFs).} solar atmosphere model; we adopt these as the best available estimates for our lines. In all cases the NLTE effects are very minor ($<$$0.03$\,dex) and we expect them to be smaller still in disk-centre intensity. We note that quantum mechanical calculations for the rate coefficients of inelastic Al+H collisions have appeared very recently (Belyaev \cite{belyaev_al}), but in view of the small importance of departures from LTE for \ali these should not modify our results significantly.

\subsection{Silicon}
Like all elements with relatively low ionisation energy, silicon is essentially \siii in the solar photosphere, but very few good \siii lines are available. We only kept one rather high excitation \siii line. We also retained a sample of nine \sii lines (Table~\ref{table:lines}).  Modern damping calculations are available for all our Si lines 
(e.g. Barklem \cite{bark2}) except the three longest wavelength lines; of these only the 703.4\,nm line has some sensitivity to the broadening, which, as mentioned above, we adopt as Uns\"old (\cite{unsold}) with an enhancement factor of two whenever more accurate broadening data are lacking.

We have derived accurate $gf$-values for the \sii lines of Table~\ref{table:lines} from the relative transition probabilities of Garz (\cite{garz}), normalised to an absolute scale with the highly accurate, laser-induced fluorescence (LIF) lifetimes of the $4s\ ^3P_{0,1,2}$ levels measured by O'Brian \& Lawler (\cite{obr1}, \cite{obr2}). We note that many more good \sii lines are available in the near infra-red, but unfortunately only low-accuracy theoretical $gf$-values are available for these lines.  We had to discard these lines for that reason, and also because they show rather large but uncertain NLTE effects (Shi et al.\ \cite{shi2}).

For the $gf$ value of our lone \siii line, we choose to use a mean of the experimental values of Schulz-Gulde (\cite{schu}), Blanco et al.\ (\cite{blan}) and Matheron et al.\ (\cite{math}).  These data were obtained by different techniques but agree quite well, resulting in an uncertainty of $\pm$0.02\,dex on the mean value.

Shi et al.\ (\cite{shi2}) analysed the formation of a large number of \sii and \siii lines in the solar flux spectrum. They confirmed a previous NLTE analysis of Wedemeyer (\cite{wede}) and show that NLTE corrections are small for the high excitation \sii lines of our sample ($\approx 0.01$\,dex) and negligible for our \siii line. We have adopted the NLTE corrections of Shi et al., which however are for flux and the theoretical MAFAGS-ODF model.  These data were calculated using the Drawin recipe with $S_{\rm H}=0.1$.

\subsection{Phosphorus}
The eight solar \phosi lines we have retained are given in Table~\ref{table:lines}. Our line selection is based on the lines used by Berzinsh et al.\ (\cite{ber}); they are all weak lines in the near-IR spectrum.  Accurate quantum-mechanical broadening data exist for all of our chosen \phosi lines.

Good $gf$-values are available from Berzinsh et al., who performed accurate new measurements of lifetimes with LIF, calculated theoretical branching fractions, and then combined them to obtain $gf$-values.  Although phosphorus is entirely $^{31}$P, with spin $I=\frac12$, no HFS is observed for our adopted weak lines.

To the best of our knowledge, no NLTE analysis exists for \altphosi. However, given the weakness of all our \phosi lines, the very small NLTE effects found for weak \suli lines, and the fact that \phosi should behave similarly to \suli because their ionisation energies are nearly identical, we can safely base our recommended solar abundance of P on the 3D LTE result.

\subsection{Sulphur}
Our sample of \suli lines (Table~\ref{table:lines}) consists of five weak lines and the well known, somewhat stronger, IR triplet at 1045\,nm. Modern damping constants exist for the IR lines, but not for the three lines in the visible part of the spectrum (e.g. Barklem \cite{bark2}).

For the IR triplet, very accurate $gf$-values have been measured by Zerne et al.\ (\cite{zern}). For the other lines we only have theoretical values, the most recent ones from Froese-Fischer \& Tachiev (\cite{FF}), Zatsarinny \& Bartschat (\cite{zats}) and Deb \& Hibbert (\cite{deb}). We adopted mean $gf$-values between as many of these three sources as available for each line; the level of agreement between the three sources suggests that uncertainties of these log $gf$-values are of order $\pm 0.05$\,dex.

For all the models we consider in this paper, we have used the ATLAS9-based (Kurucz \cite{atlas9}) NLTE corrections of Takeda et al.\ (\cite{tak1}), computed for our chosen lines in disk-centre intensity.  The six weak \suli lines of our sample have very small NLTE effects, whereas those for the IR triplet are rather large.  We chose $S_{\rm H}=0.4$, as this leads to the best agreement between the two groups of lines when the offsets are applied to the 3D LTE results.

\subsection{Potassium}
Because of its very low ionisation energy, K is essentially all \kii in the solar photosphere.  However, the only useful lines in the solar spectrum for abundance derivation are \altki.  We retained five weak \ki lines (Table~\ref{table:lines}) after examining the recent work of Zhang et al.\ (\cite{zhan}).  Modern data only exist for the impact of hydrogen collisions on the broadening of the two lines with the longest wavelengths.  In AGSS09 we also included the 769.9\,nm line; here we have decided to exclude this line because it is very strong, and extremely sensitive to departures from LTE.

We adopted $gf$-values from a number of sources.  Where possible we used experimental results: either the intracavity laser measurement of Gamalii (\cite{Gamalii97}, as recommended by Sansonetti \cite{Sansonetti08}), or, as per Morton (\cite{Morton03}), the data obtained by the hook method by Shabanova \& Khlyustalov (\cite{Shabanova85}), renormalised to the mean of accurate lifetimes measured by Volz \& Schmoranzer (\cite{Volz96}) and Wang et al.\ (\cite{Wang97}).  For one line (580.2\,nm), we used a theoretical $gf$-value calculated by the Opacity Project under the assumption of \textit{LS} coupling, provided to Zhang et al (\cite{zhan}) by Keith Butler.  The experimental $gf$-values should be reasonably accurate, the theoretical ones from the Opacity Project less so.  The experimental value for the 693.9\,nm line from Gamalii (\cite{Gamalii97}) has a stated accuracy of only 40\%, but the internal abundance scatter of both our results and those of Zhang et al.\ (\cite{zhan}), when this $gf$ value is employed instead of the alternative theoretical one from the Opacity Project, suggests that this is likely an underestimated uncertainty.

\ki lines show some HFS, as potassium consists of 93.3\% $^{39}$K, which has nuclear spin $I=\frac32$.  Good HFS measurements exist for all levels of the lines we derive abundances from.  The most accurate data come from Bloom \& Carr (\cite{Bloom60}), Dahmen \& Penselin (\cite{Dahmen67}), Svanberg (\cite{Svanberg71}), Gupta et al.\ (\cite{Gupta73}), Belin et al.\ (\cite{Belin75}), Sieradzan et al.\ (\cite{Sieradzan97}) and Falke et al.\ (\cite{Falke06}).  We prefer the results of Falke et al.\ (\cite{Falke06}) to those of Das \& Natarajan (\cite{Das08}) for the $4p$\ $^2P_{1/2,3/2}$ levels, as data from the latter appear not to agree with other measurements (e.g.\ Refs.\ 21 and 22 in that paper).  In AGSS09 the HFS constants for the upper and lower levels of \ki were erroneously reversed; this has been remedied in the present analysis.

We computed flux and intensity NLTE corrections for the ATLAS9 model following Takeda et al.\ (\cite{tak2}), and applied the disk-centre corrections to LTE results from each of the atmospheric models that we consider here. The flux values are in very good agreement with the recent NLTE results of Zhang et al.\ (\cite{zhan}). We computed corrections for different values of $S_{\rm H}$, finally adopting $S_{\rm H}=0.4$ because it produces the best agreement between different lines (Table~\ref{table:lines}).

\subsection{Calcium}
Ca has a relatively low ionisation potential, so exists mostly as \caii in the solar photosphere.  We examined the lines used in the very complete and detailed NLTE analysis of Mashonkina et al.\ (\cite{mash1}), ultimately retaining eleven \cai lines and five \caii lines (Table~\ref{table:lines}). The \cai lines are somewhat stronger than the \caii lines. We do not include the inter-combination line at 657.28\,nm in our sample, despite its inclusion in some past analyses, because it is formed in the outer wing of the H${\alpha}$ line. The permitted \caii lines have high excitation potentials.  Except for the 824.8\,nm line, which has a high transition probability, they therefore form relatively low in the atmosphere and are hence less influenced by atmospheric inhomogeneities, and possibly also by NLTE effects (although this can be offset by their comparatively small level populations).  The forbidden line at 732.3\,nm is formed in perfect LTE because it is a weak transition from the ground state of the dominant ionisation stage.

Modern damping constants are available for all the \cai lines we use except 451.2\,nm and 586.8\,nm.  These are both rather weak lines, so this should have no significant impact on the derived abundance.  No such data exist for \altcaii.

Highly accurate experimental $gf$-values are available for the \cai lines from Smith \& Raggett (\cite{smit}) and Smith (\cite{Smith88}). For \altcaii, no very accurate $gf$-values exist for the permitted lines; we relied on the choices discussed by Mashonkina et al.\ (\cite{mash1}), with data ultimately coming from Opacity Project calculations under the assumption of $LS$ coupling.  We chose to use the $gf$ value of Mel\'endez et al.\ (\cite{melen}) for the forbidden line, taking the mean value from 4 recent theoretical results reported by those authors; the uncertainty should be of order 0.03\,dex.

We have computed the NLTE abundance correction for Ca using a new Ca model atom (Lind et al. \cite{lind_li6}, and references therein).  We calculated disk-centre intensity and flux spectra for the HM, \marcs\ and \oneDAV\ model atmosphere.  As with other elements, in the absence of dedicated 3D NLTE calculations we adopt the \oneDAV\ calculations as a proxy for the full 3D results. No quantum mechanical calculations for inelastic Ca+H collisions have yet been published, so we adopt $S_{\rm H}=0.1$, as for \altmgii.

\begin{figure*}
\centering
\begin{minipage}[t]{0.37\textwidth}
\centering
\includegraphics[width=\linewidth]{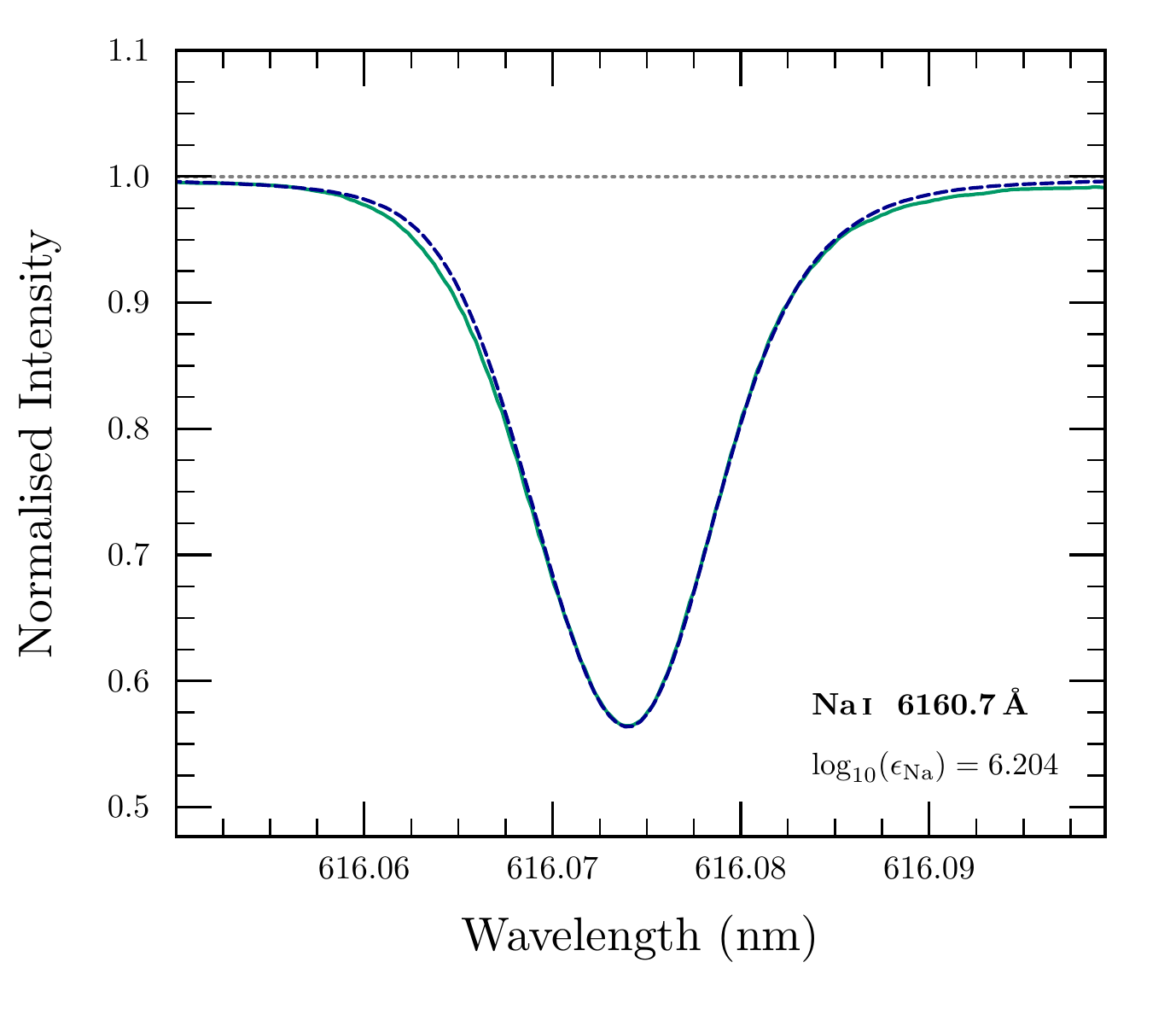}
\includegraphics[width=\linewidth]{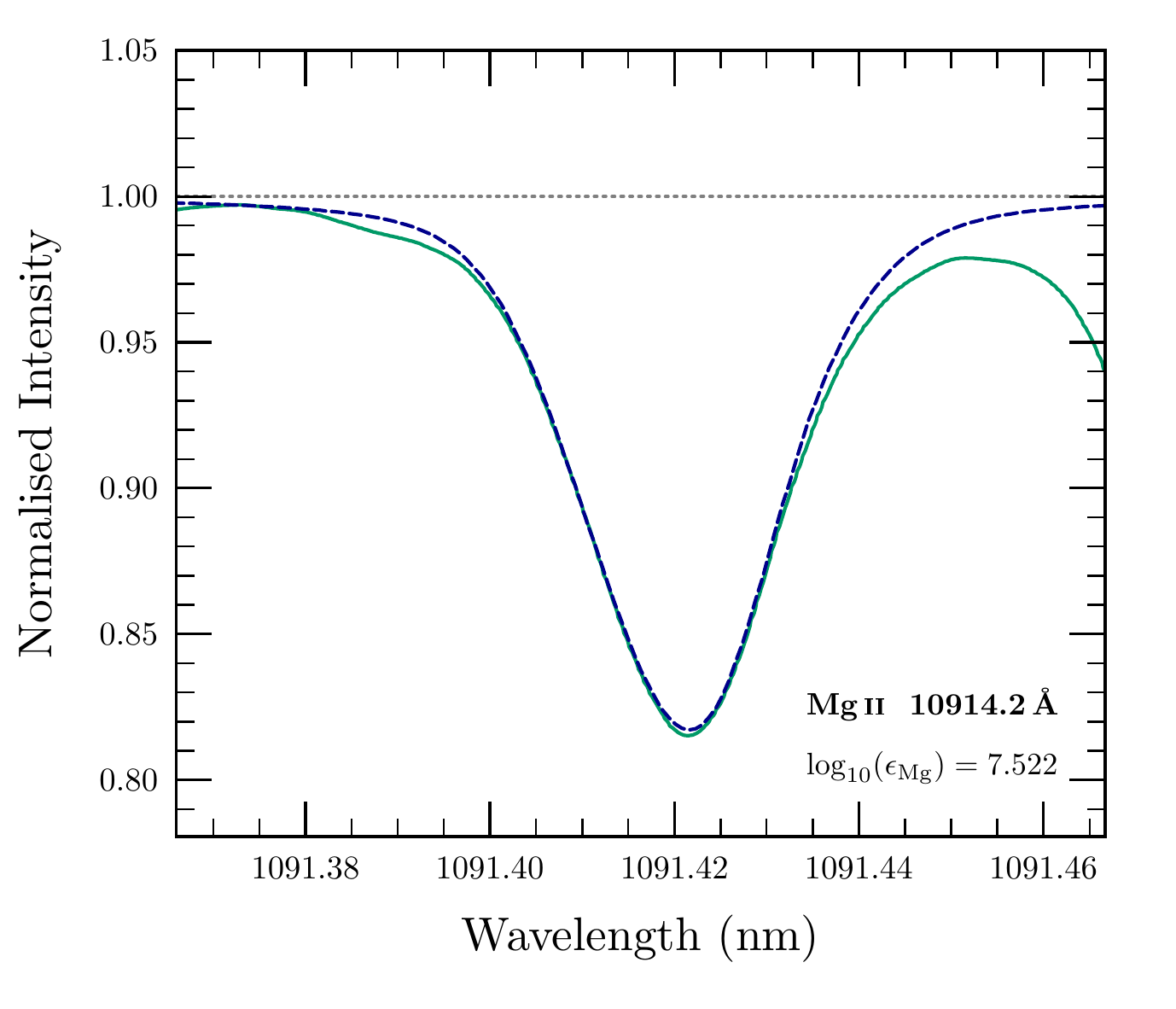}
\includegraphics[width=\linewidth]{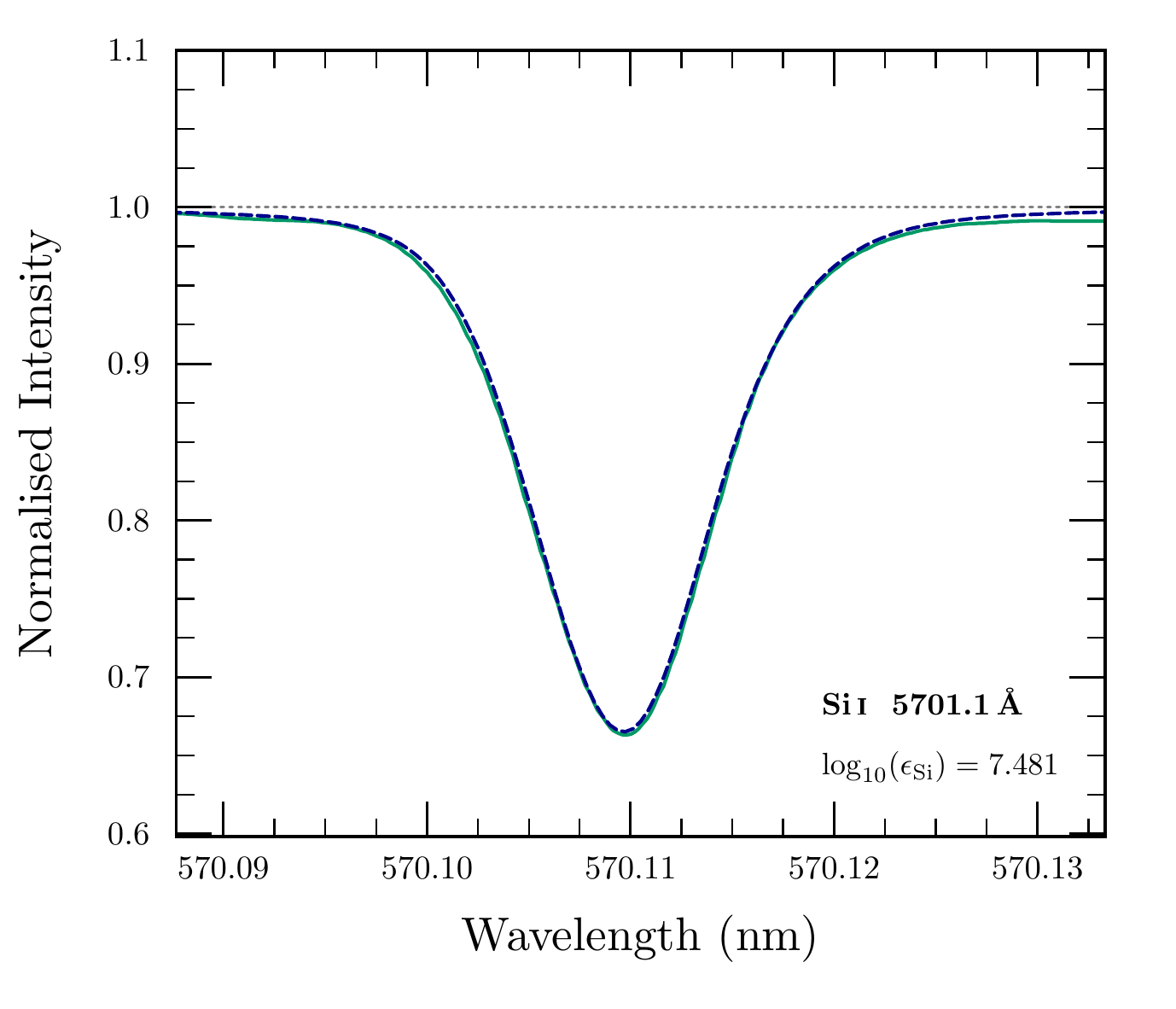}
\includegraphics[width=\linewidth]{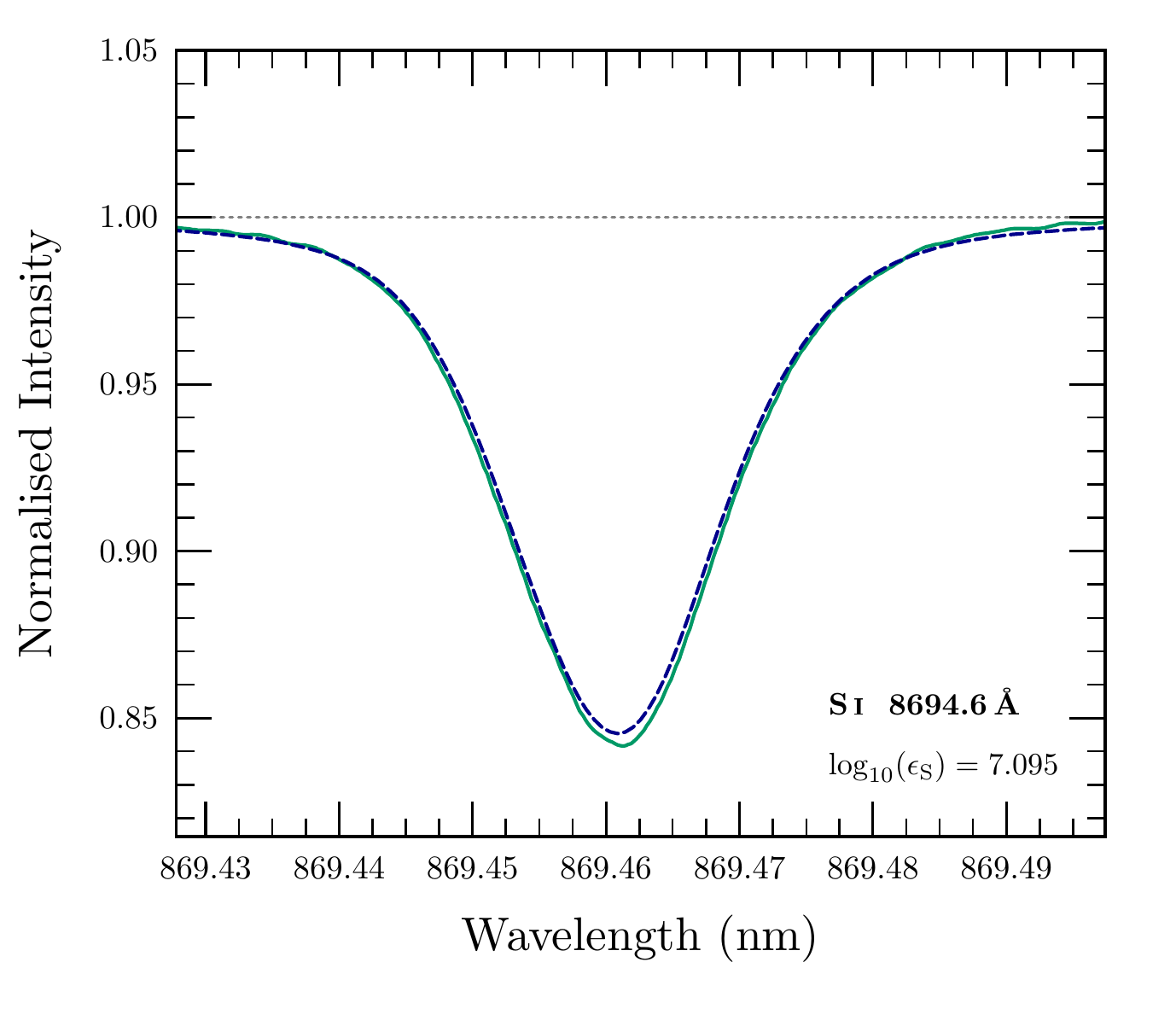}
\end{minipage}
\hspace{0.05\textwidth}
\begin{minipage}[t]{0.37\textwidth}
\centering
\includegraphics[width=\linewidth]{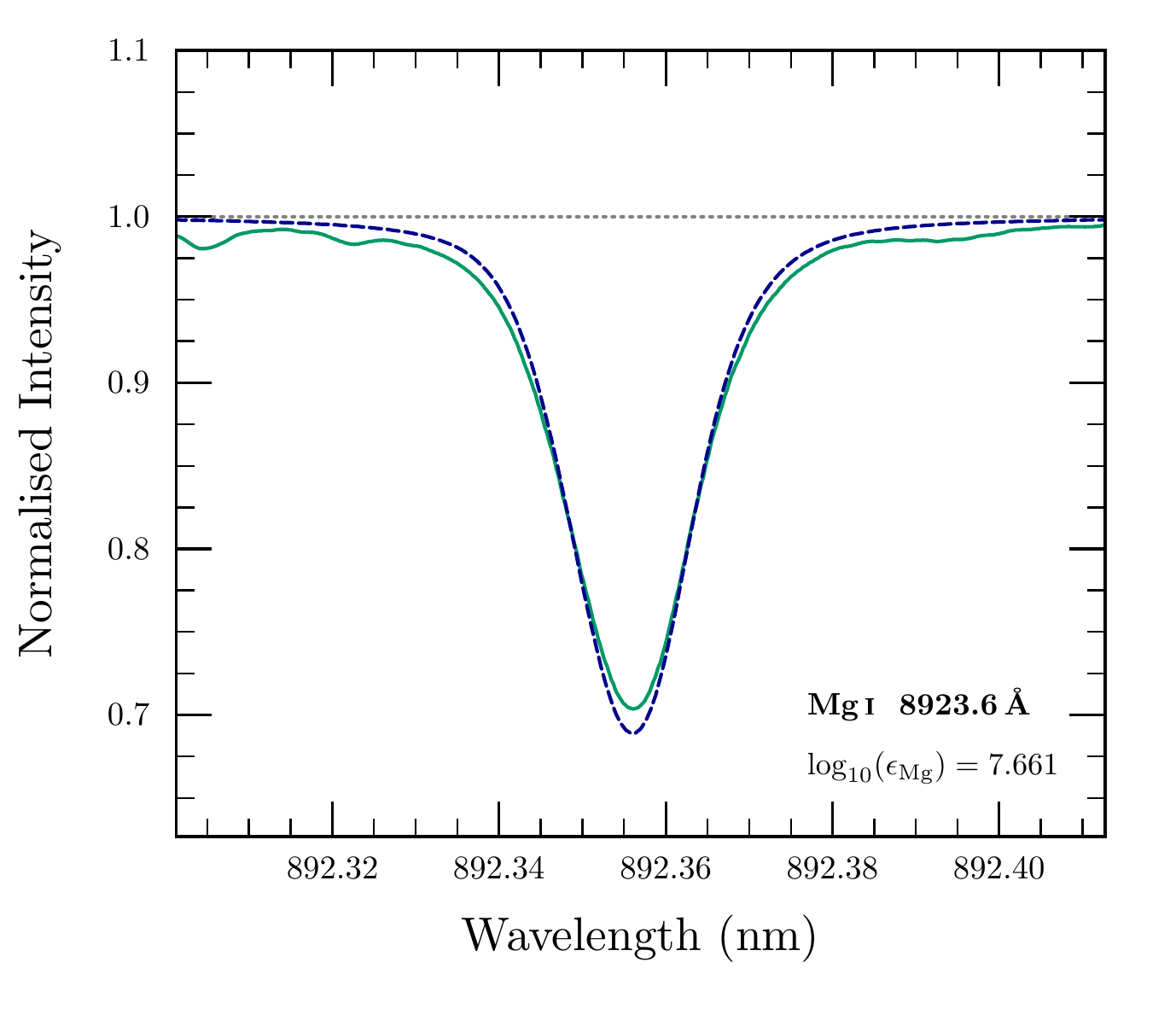}
\includegraphics[width=\linewidth]{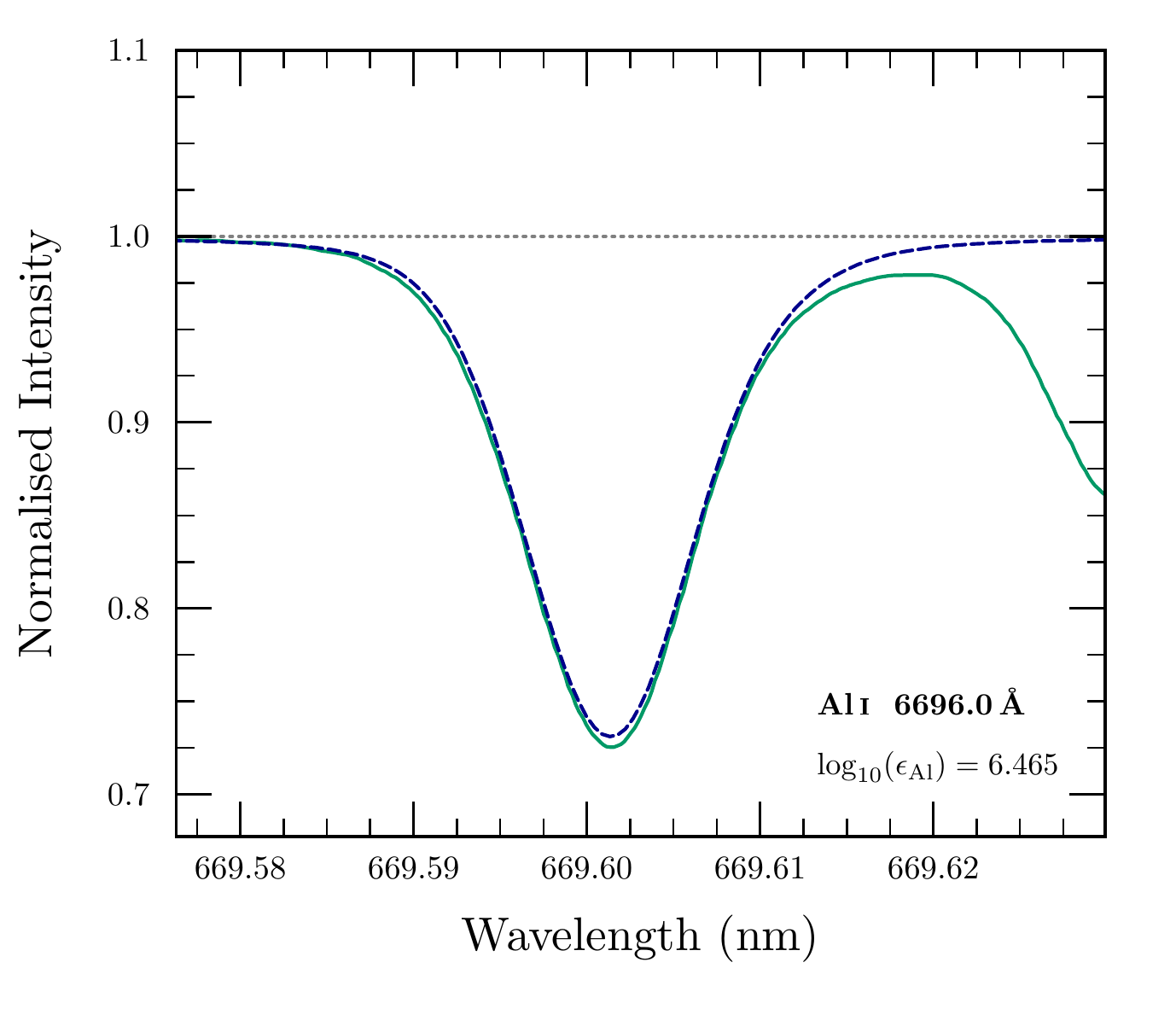}
\includegraphics[width=\linewidth]{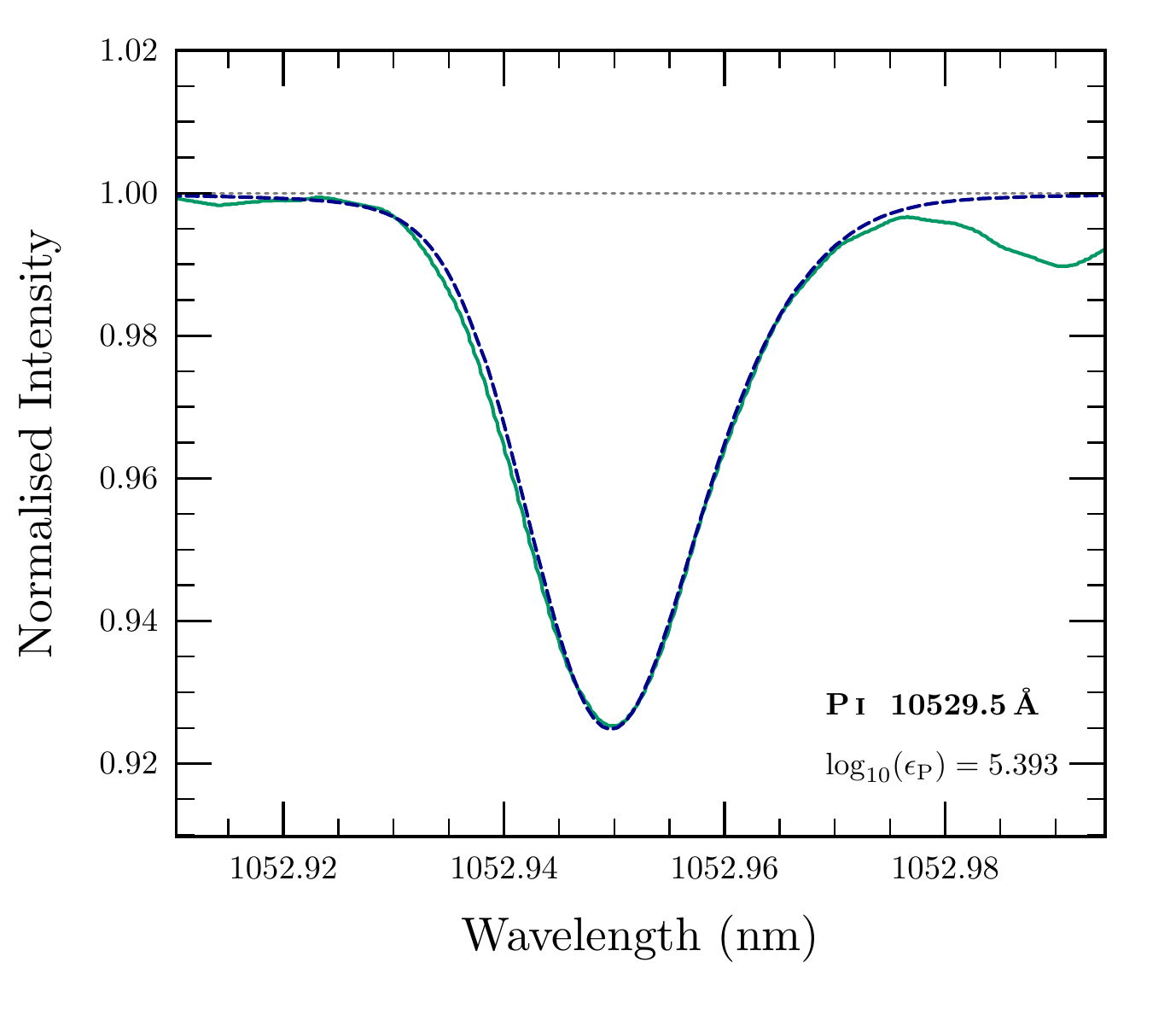}
\includegraphics[width=\linewidth]{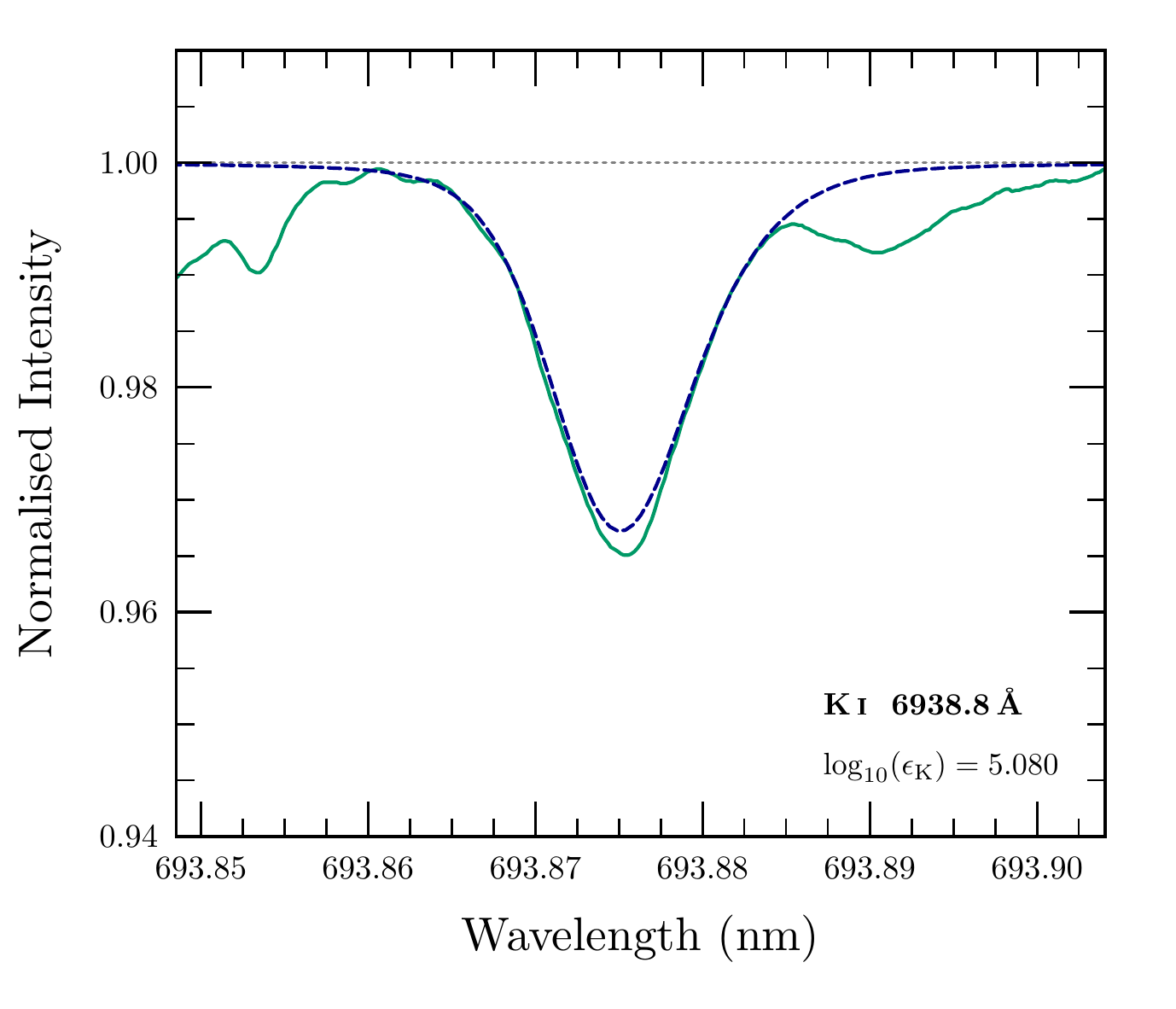}
\end{minipage}
\caption{Example spatially and temporally averaged, disk-centre synthesised \altnai, \altmgi, \altmgii, \altali, \altsii, \altphosi, \suli and \ki line profiles (blue dashed), shown in comparison to the observed Kitt Peak FTS profile (solid green).  We removed the solar gravitational redshift from the FTS spectrum, convolved the synthesised profile with an instrumental sinc function and fitted it in abundance.  In some cases the observed spectra have been adjusted slightly in wavelength and continuum placements relative to the published solar atlas. The 3D line profiles shown are the 3D LTE results, whereas the quoted abundance in each panel includes the NLTE abundance correction from Table \ref{table:lines}.}
\label{fig:profiles}
\end{figure*}

\begin{figure*}
\centering
\begin{minipage}[t]{0.37\textwidth}
\centering
\includegraphics[width=\linewidth]{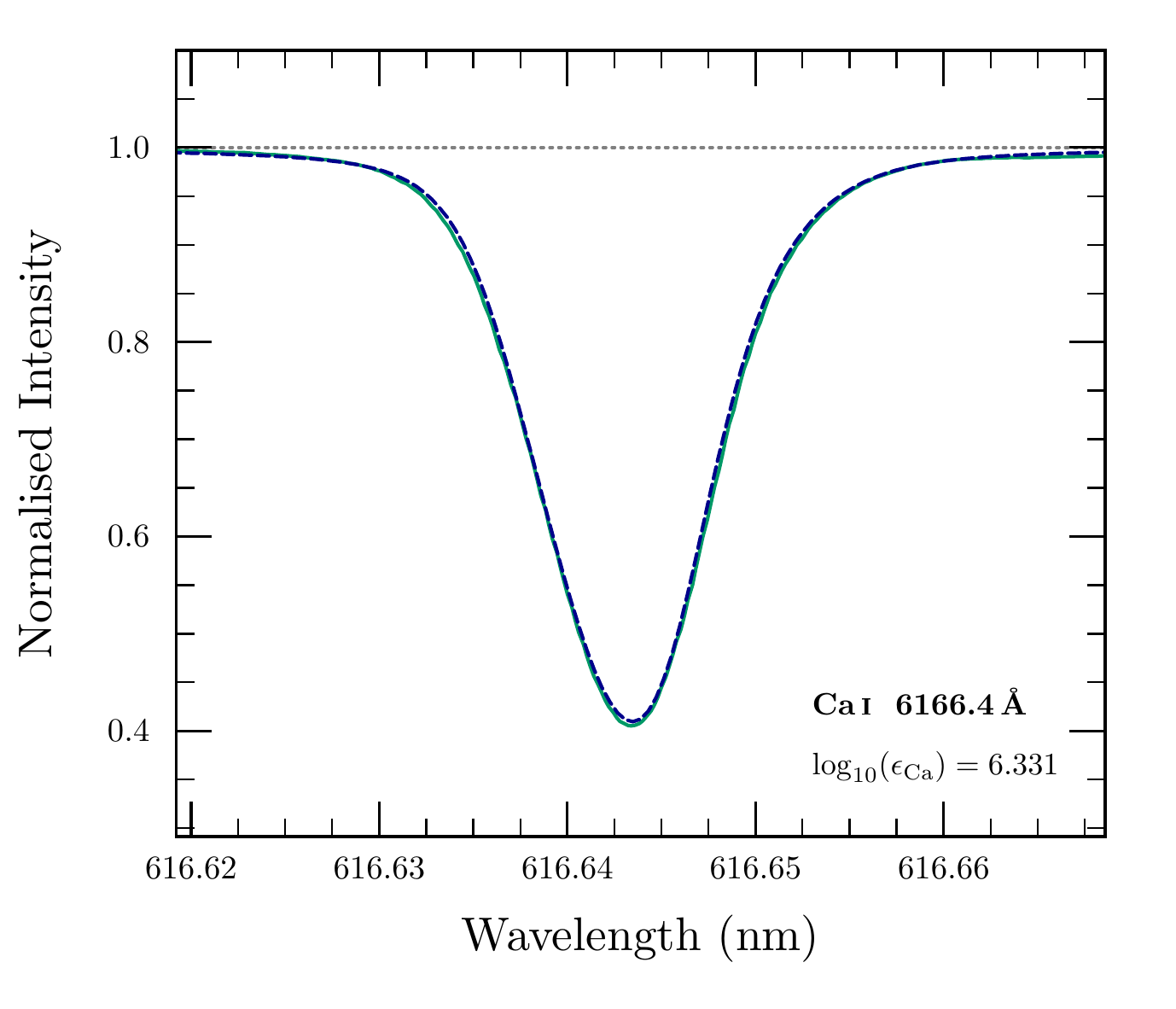}
\end{minipage}
\hspace{0.05\textwidth}
\begin{minipage}[t]{0.37\textwidth}
\centering
\includegraphics[width=\linewidth]{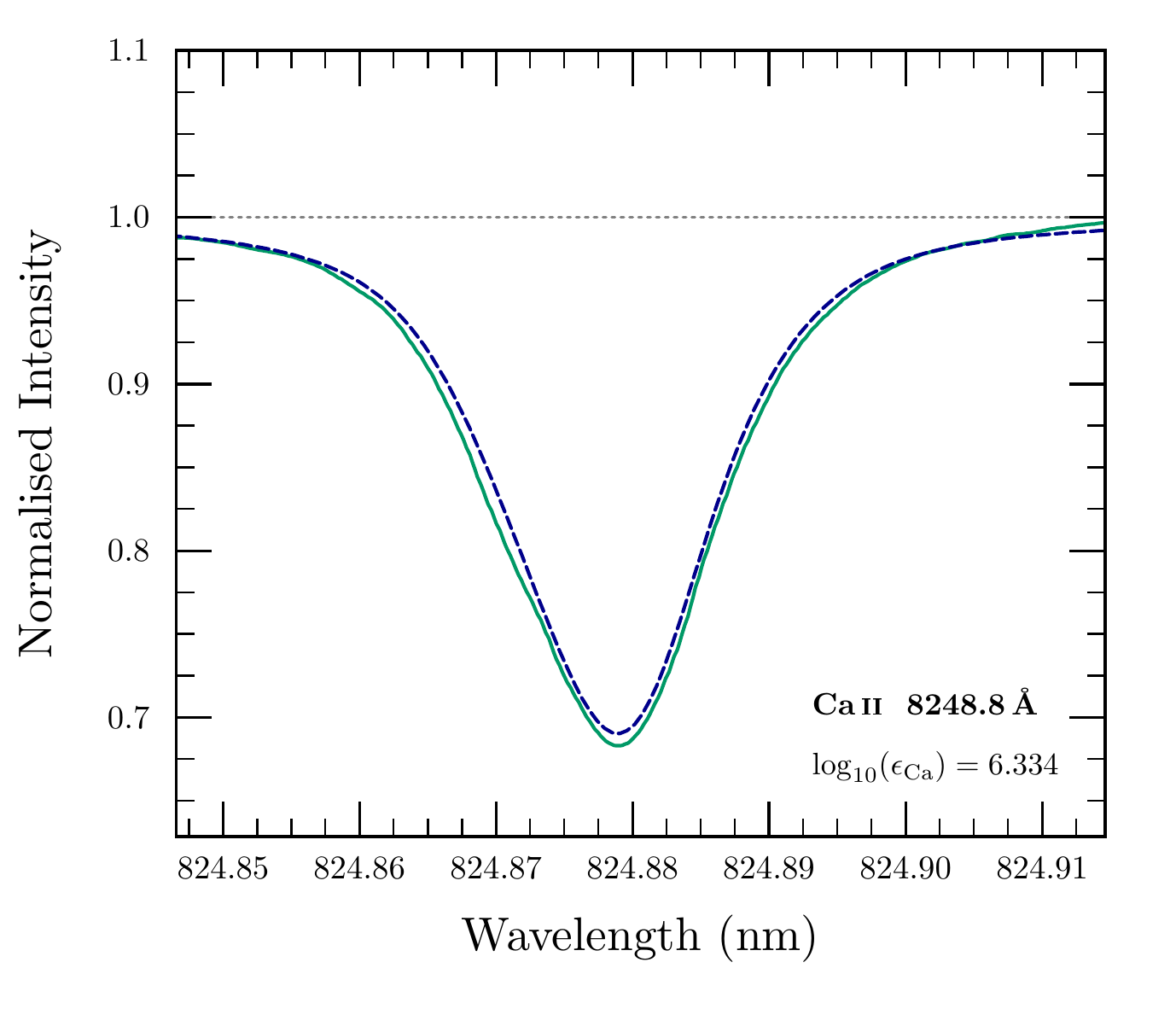}
\end{minipage}
\caption{As per Fig.~\protect\ref{fig:profiles}, but for \cai and \caii.}
\label{fig:profiles2}
\end{figure*}

\section{Derived solar elemental abundances}
\label{s:results}

For the abundance determinations in this series of papers, we use equivalent widths measured on disc-centre solar intensity spectra (cf. Sect.\ \ref{s:abundancecalculation}). However, we also compared all the observed line profiles to the profiles predicted by the 3D model. In Figs.~\ref{fig:profiles} and \ref{fig:profiles2} we show a series of representative line profiles for the elements that we consider in this paper. Agreement between predicted and observed profiles is typically very good, showing that the 3D model is without doubt highly realistic.  No 1D calculation can come close to the 3D result in terms of resemblance to the observed spectrum, even with the inclusion of \textit{ad hoc} free parameters for micro- and macroturbulence.  This is no surprise, as those parameters are simply designed to mimic effects of three-dimensional fluid flow, like line broadening, shifts and asymmetries arising from convective motions and oscillations in the atmosphere (e.g. Asplund et al. \cite{asp1}).  Despite the excellent agreement, some small discrepancies between the predicted 3D profiles and observed spectrum often remain.  These are most typically in the line cores, and can often be attributed to NLTE effects absent in the 3D radiative transfer calculation (although we correct the actual abundances for NLTE effects \textit{a posteriori} whenever possible), or to missing broadening data (as can be clearly seen in the example of the \mgi line in Fig.~\ref{fig:profiles}).

\subsection{Sodium}
We find a mean 3D+NLTE abundance of Na of $\log\epsilon_{\mathrm{Na}}= 6.21 \pm 0.01$ (s.d.). The results obtained with the various 1D models described in Sect. \ref{s:atmospheres} are given in Table~\ref{table:summary}. If the total uncertainty is estimated as explained in Sect.~\ref{s:errors}, the solar Na abundance becomes $\log\epsilon_{\mathrm{Na}} = 6.21 \pm 0.04$ ($\pm$$<$0.01 stat, $\pm$0.04 sys).

As expected, with 1D semi-empirical models the inferred Na abundance is somewhat higher (Table~\ref{table:summary}) due mainly to the shallower temperature gradient ($\approx$$0.05$\,dex) and but also partly in view of the absence of atmospheric inhomogeneities ($\approx$$0.02$\,dex). As is almost always the case for both neutral and ionised species, \marcs\ returns the lowest Na abundance, mainly because its temperature stratification is too steep, as seen by comparison with the observed solar centre-to-limb variation (Pereira et al. \cite{pereira_models}).

\begin{figure*}
\centering
\begin{minipage}[t]{0.4\textwidth}
\centering
\includegraphics[width=\linewidth]{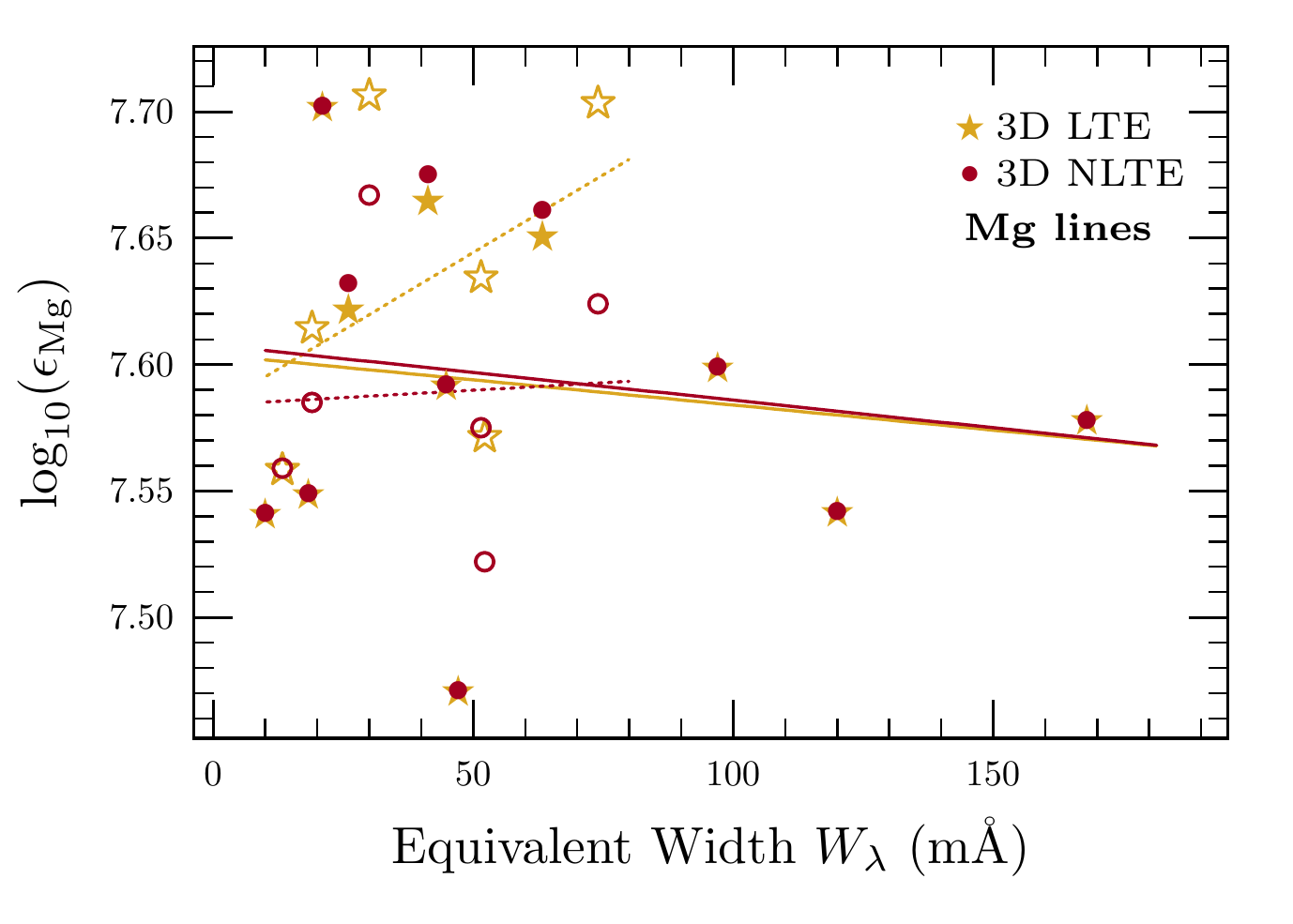}
\end{minipage}
\hspace{0.05\textwidth}
\begin{minipage}[t]{0.4\textwidth}
\centering
\includegraphics[width=\linewidth]{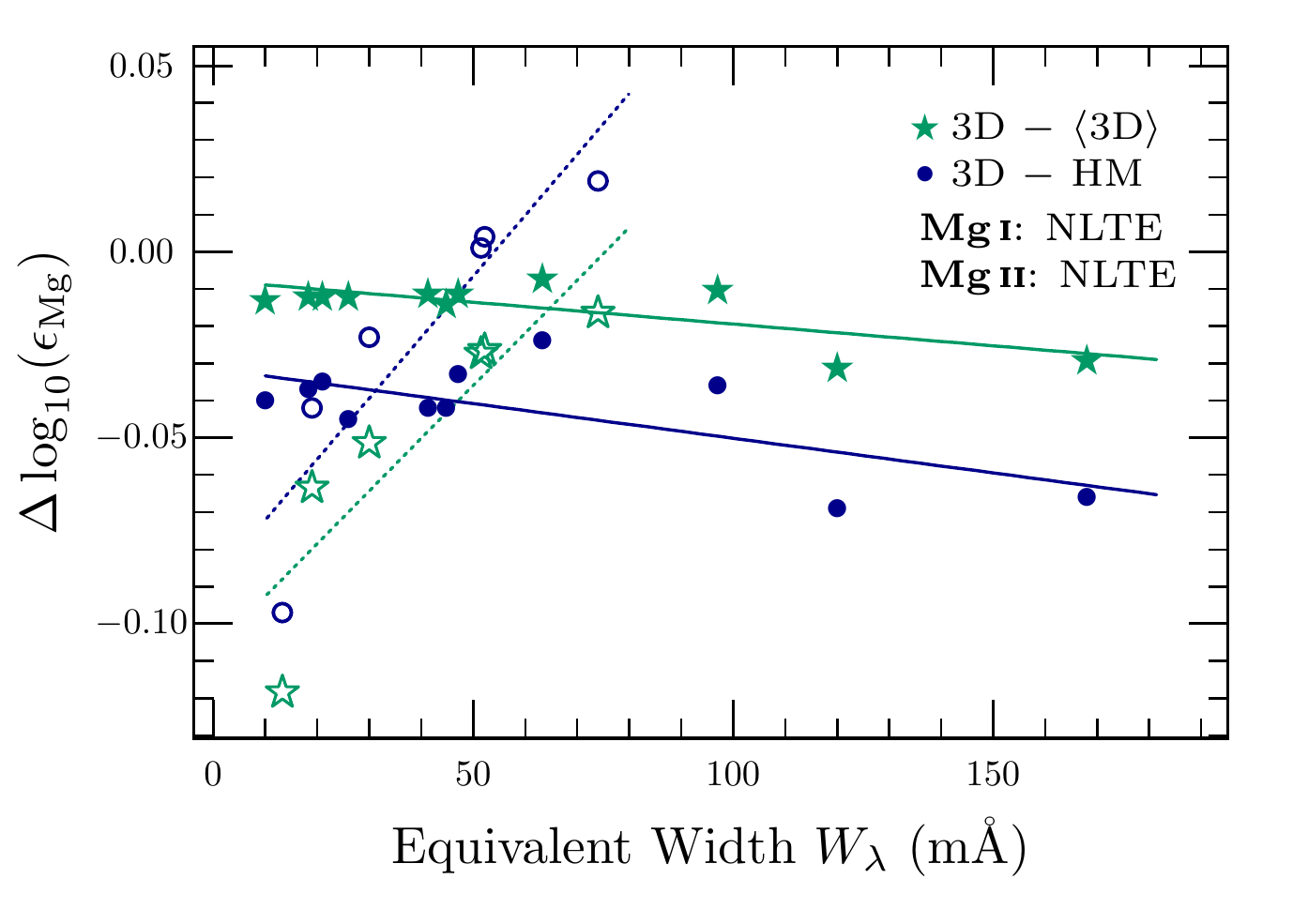}
\end{minipage}
\caption{\textit{Left}: Mg abundances derived from \mgi and \mgii lines with the 3D model, shown as a function of equivalent width.
\textit{Right}: Line-to-line differences between Mg abundances obtained with the 3D and \oneDAV\ models, and between those obtained with the 3D and HM models.  Filled symbols and solid trendlines indicate lines of the neutral species (\altmgi), whereas open symbols and dotted lines indicate singly-ionised (\altmgii) lines.  Trendlines give equal weight to each line (unlike our mean abundances, where we give larger weights to higher quality lines).}
\label{fig:mg}
\end{figure*}

\subsection{Magnesium}
Line-to-line behaviour of our Mg abundances with equivalent width is shown in Fig.\ \ref{fig:mg}.  The mean 3D+NLTE abundance of Mg from our sample of 11 \mgi lines is $\log\epsilon_{\mathrm{Mg}} = 7.60 \pm 0.07$ (s.d.), and from the 6 \mgii lines, $\log\epsilon_{\mathrm{Mg}} = 7.58 \pm 0.05$ (s.d.). The mean solar Mg abundance, across all our chosen \mgi and \mgii lines, becomes $\log\epsilon_{\mathrm{Mg}} = 7.59 \pm 0.04$ ($\pm$0.01 stat, $\pm$0.03 sys).  

Results with 1D models are given in Table~\ref{table:summary}. The 3D result is in between the \marcs-based value and those from all other 1D models. The overall impacts on the mean Mg abundance of temperature inhomogeneities and the mean temperature stratification are roughly similar, although the two species have slightly different sensitivities to the two properties.

\begin{figure*}
\centering
\begin{minipage}[t]{0.4\textwidth}
\centering
\includegraphics[width=\linewidth]{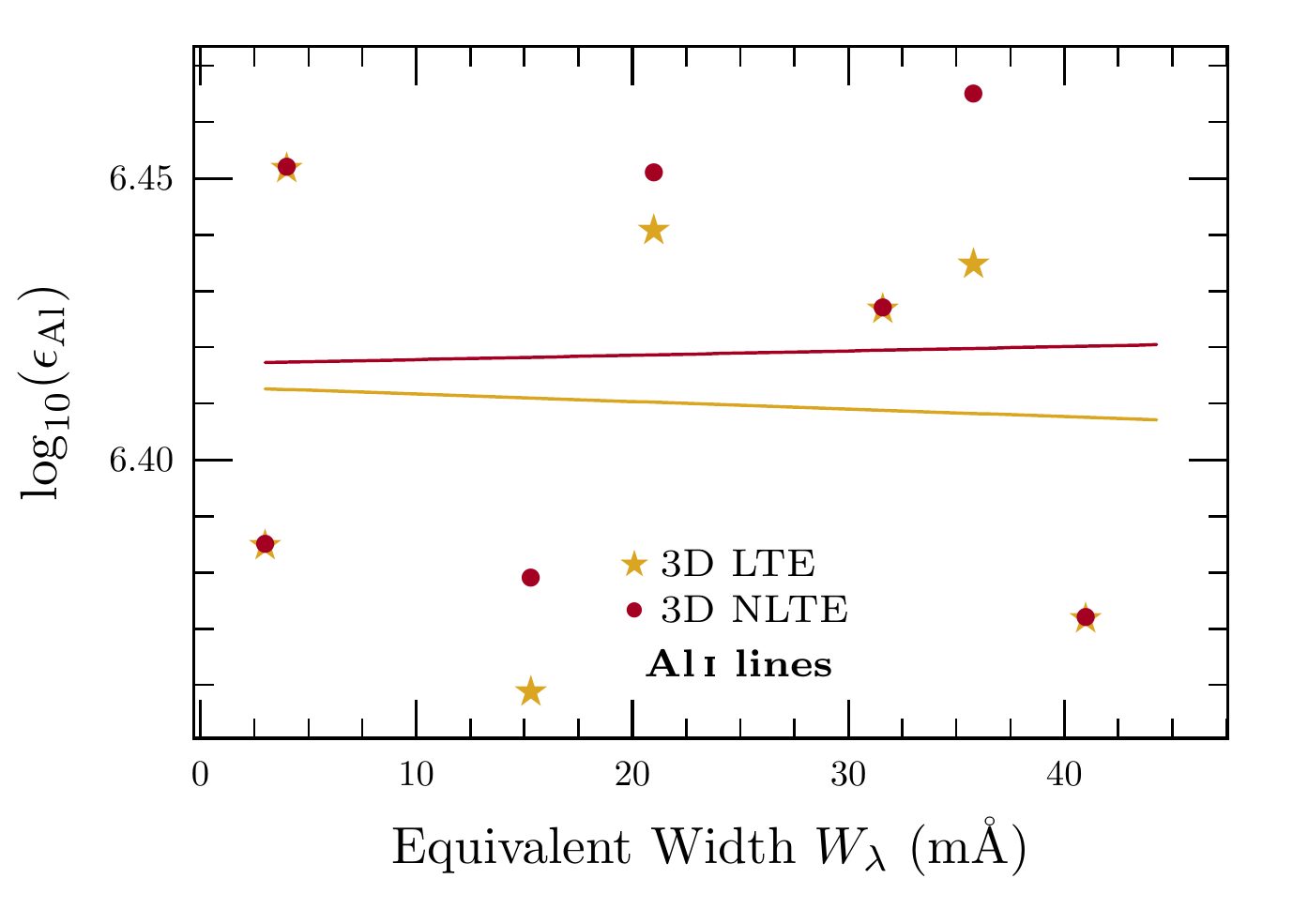}
\end{minipage}
\hspace{0.05\textwidth}
\begin{minipage}[t]{0.4\textwidth}
\centering
\includegraphics[width=\linewidth]{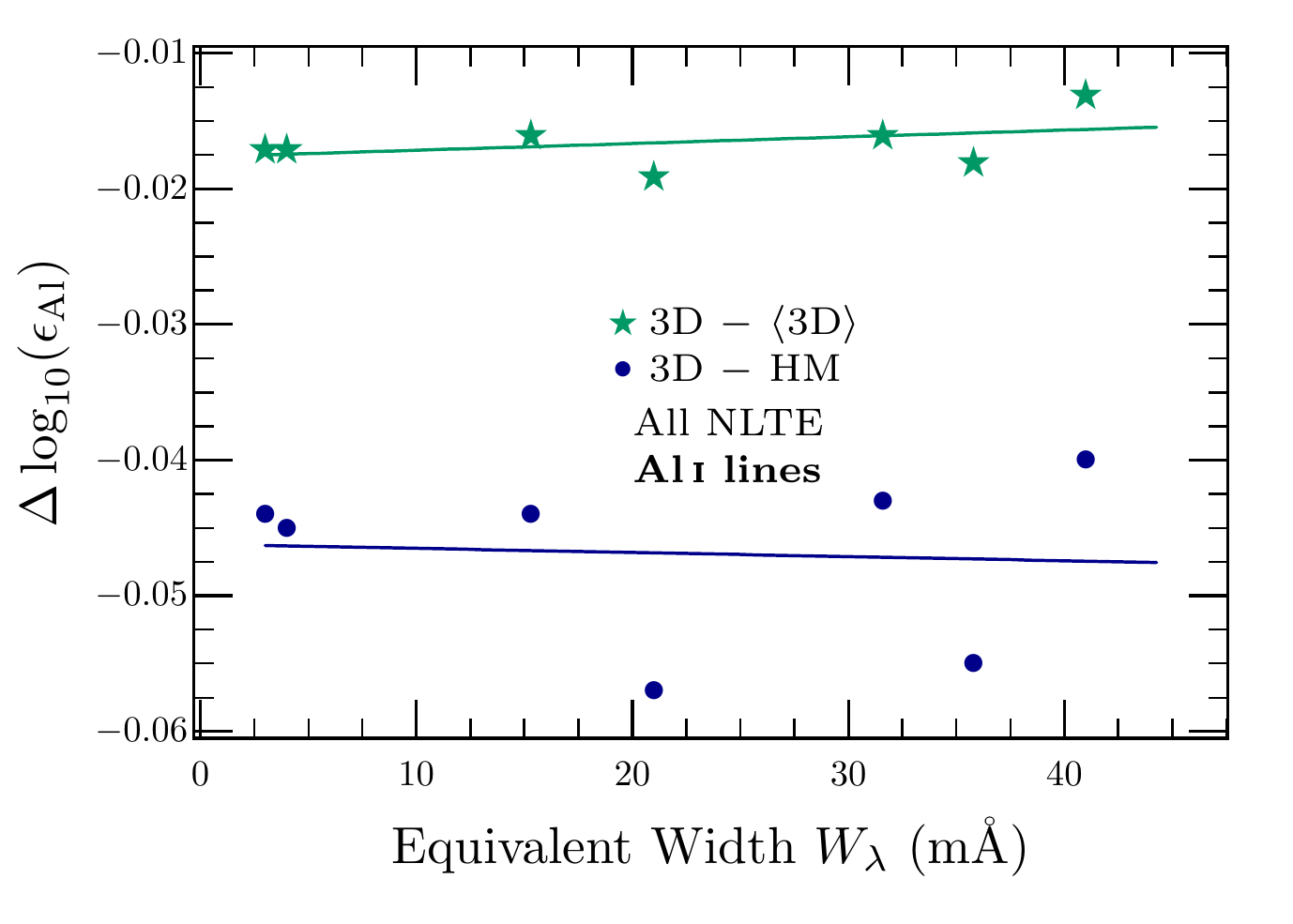}
\end{minipage}
\caption{\textit{Left}: 3D Al abundances from \ali lines, as a function of equivalent width.  \textit{Right}: Line-to-line differences between abundances obtained with the 3D and \oneDAV\ models, and between those obtained with the 3D and HM models.}
\label{fig:al}
\end{figure*}

\subsection{Aluminium}
Line-to-line behaviour of our Al abundances with equivalent width is shown in Fig.\ \ref{fig:al}.  The 3D+NLTE abundance of Al is $\log\epsilon_{\mathrm{Al}} = 6.43 \pm 0.04$ ($\pm$0.01 stat, $\pm$0.04 sys). 
Due to its high temperature sensitivity, the derived \ali abundance with the HM model is significantly larger, but the 3D-\oneDAV\ difference is rather small (Table~\ref{table:summary}). 

\begin{figure*}
\centering
\begin{minipage}[t]{0.4\textwidth}
\centering
\includegraphics[width=\linewidth]{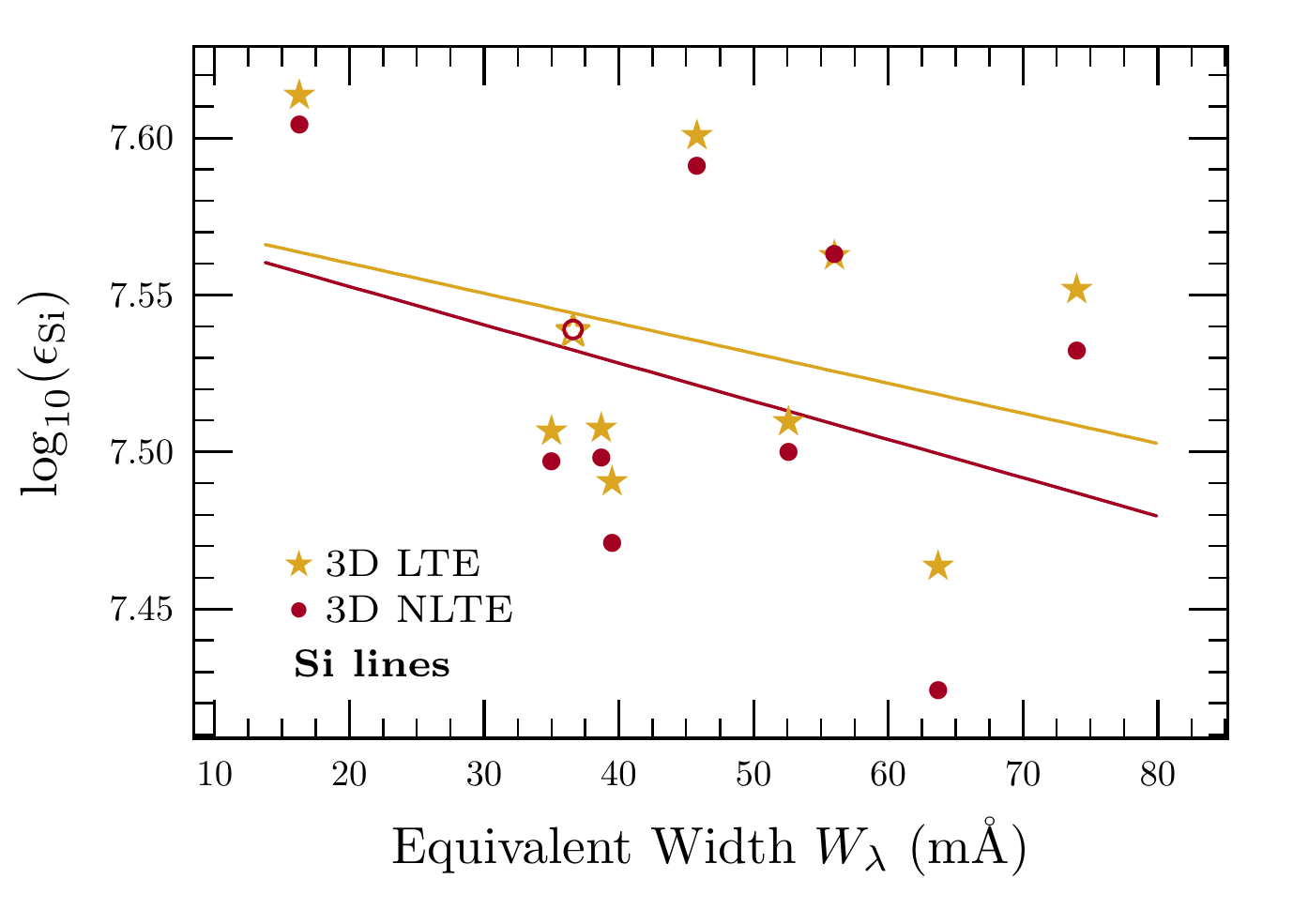}
\end{minipage}
\hspace{0.05\textwidth}
\begin{minipage}[t]{0.4\textwidth}
\centering
\includegraphics[width=\linewidth]{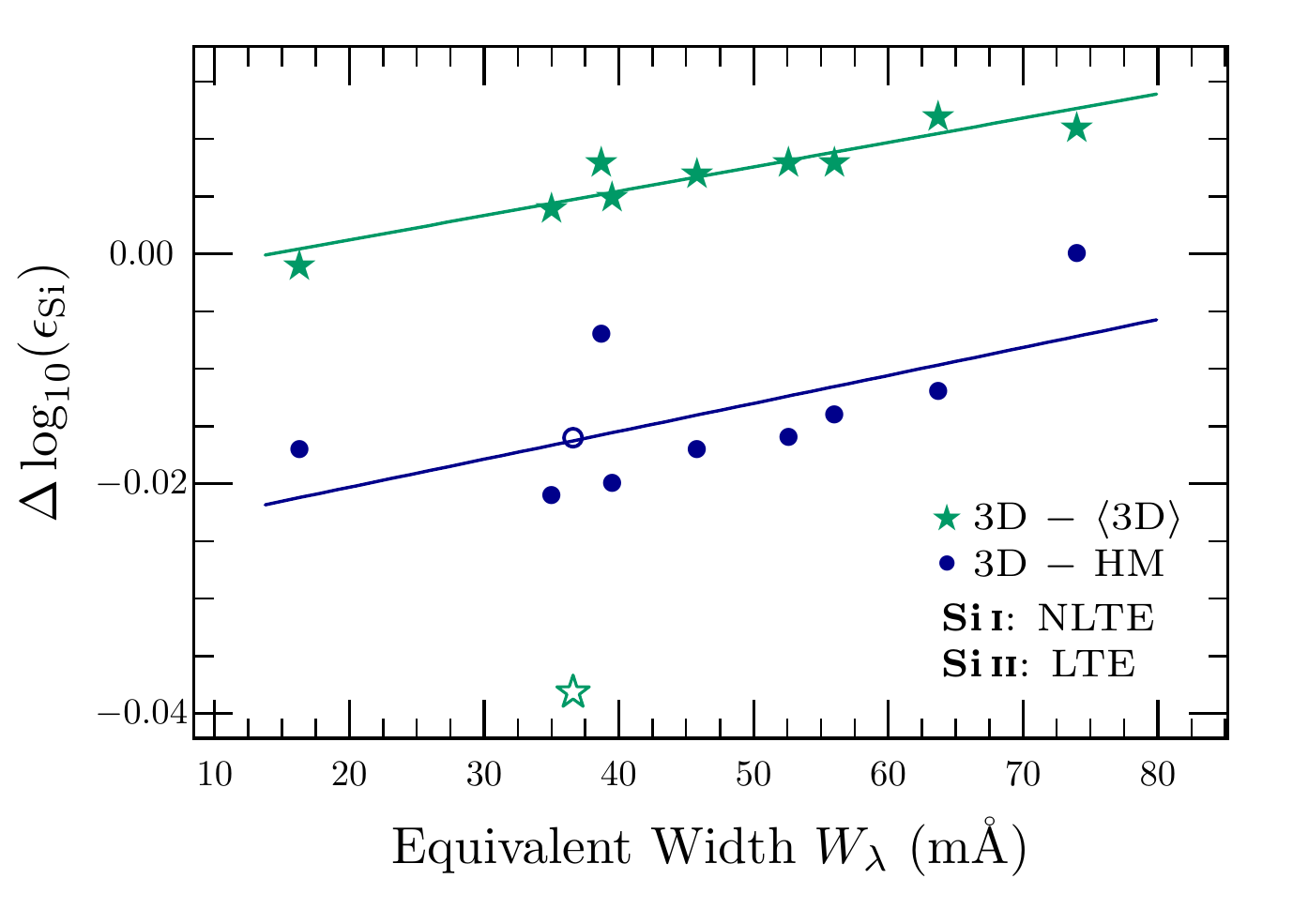}
\end{minipage}
\caption{\textit{Left}: 3D Si abundances from \sii lines (solid symbols) and the single \siii line (open symbols), as a function of equivalent width.  
\textit{Right}: Line-to-line differences between abundances obtained with the 3D and \oneDAV\ models, and between those obtained with the 3D and HM models.}
\label{fig:si}
\end{figure*}

\subsection{Silicon}
Line-to-line behaviour of our Si abundances with equivalent width is shown in Fig.\ \ref{fig:si}.  Our 3D+NLTE results are the following: $\log\epsilon_{\mathrm{Si}}= 7.51 \pm 0.05$ (s.d.; 9 \sii lines) and $\log\epsilon_{\mathrm{Si}} = 7.54$ (one \siii line).  Taking the mean of all lines, the final silicon abundance is $\log\epsilon_{\mathrm{Si}} = 7.51 \pm 0.03$ ($\pm$0.01 stat, $\pm$0.03 sys). There is a slight negative correlation between the derived \sii abundance and the line strength in the 3D case. With the exception of the low \marcs-based result, all model atmospheres return very similar abundances. 

The Si abundance obtained by Shi et al. (\cite{shi2}) was $7.52 \pm 0.06$, from a series of rather strong \sii lines together with two \siii lines observed in solar flux spectra, and interpreted with a 1D theoretical model in NLTE.  This is in very good agreement with our 3D+NLTE result.  It is significantly larger than the value we found with the theoretical \marcs\ model though, which should be quite similar to the one employed by Shi et al. (\cite{shi2}); this can be explained by the fact that Shi et al.\ used the original oscillator strengths of Garz (\cite{garz}) without renormalising them to accurate lifetimes as we did.

\subsection{Phosphorus}
Our raw 3D result is $\log\epsilon_{\mathrm{P}} = 5.41 \pm 0.03$ (s.d.).  Including the full error budget, the final result is $\log\epsilon_{\mathrm{P}} = 5.41 \pm 0.03$ ($\pm$0.01 stat, $\pm$0.03 sys).
As expected for such high-excitation lines, the \phosi\ results are hardly sensitive to the adopted model atmosphere, be it 1D or 3D (Table~\ref{table:summary}).
The \phosi\ lines span too small a range in equivalent width and excitation potential to trace any trends with any of the 3D or 1D analyses.

Caffau et al.\ (\cite{caf1}) also analysed five of our IR lines with their own 3D model: 1051.1, 1052.9, 1058.1, 1059.6 and 1068.1\,nm, adopting the same $gf$-values as we do. After discarding the 1068.1\,nm line because it implied an abundance 0.1\,dex lower than the rest, their recommended value is slightly higher than ours:  $\log\epsilon_{\mathrm{P}} = 5.46 \pm 0.04$ (s.d.). In contrast, our study contains another three lines and results in no such outliers, as evident from our very small line-to-line scatter. It is not obvious what caused Caffau et al.'s discrepant result for the 1068.1\,nm line, as their equivalent width for disk-centre intensity is very similar to ours.  We note that their derived abundance for the 1051.1\,nm line is also surprisingly large, in spite of the fact that their adopted equivalent width is identical to our measured value.  These are all very weak lines, with similar excitation potentials; in our case, as expected, the measured line strengths essentially perfectly reflect the differences in transition probabilities of the lines, and the variations in strength are consistent with the inferred abundance scatter.

\begin{figure*}
\centering
\begin{minipage}[t]{0.4\textwidth}
\centering
\includegraphics[width=\linewidth]{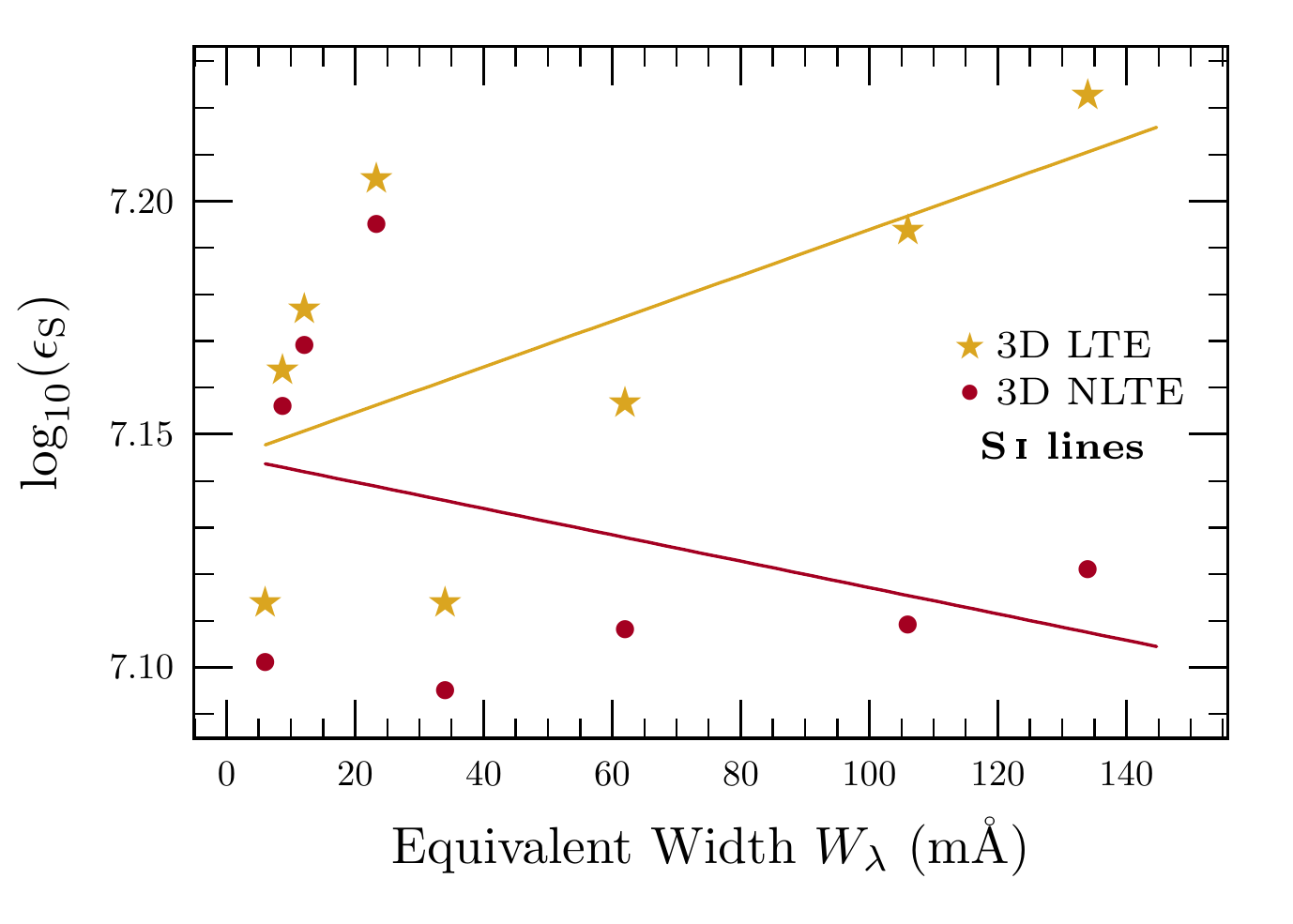}
\end{minipage}
\hspace{0.05\textwidth}
\begin{minipage}[t]{0.4\textwidth}
\centering
\includegraphics[width=\linewidth]{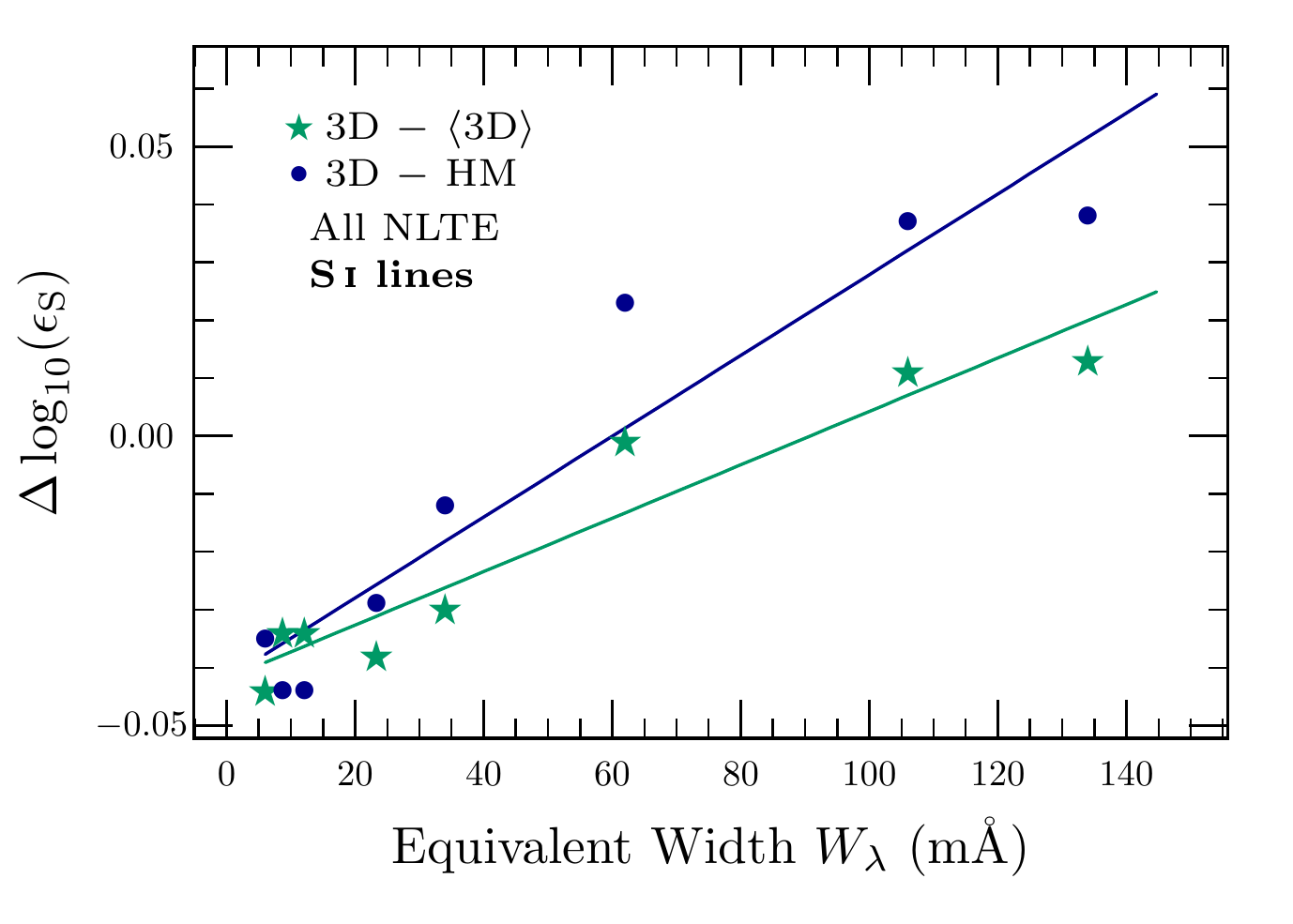}
\end{minipage}
\caption{\textit{Left}: 3D S abundances from \suli lines, as a function of equivalent width. 
\textit{Right}: Line-to-line differences between abundances obtained with the 3D and \oneDAV\ models, and between those obtained with the 3D and HM models.}
\label{fig:s}
\end{figure*}

\begin{figure*}
\centering
\begin{minipage}[t]{0.4\textwidth}
\centering
\includegraphics[width=\linewidth]{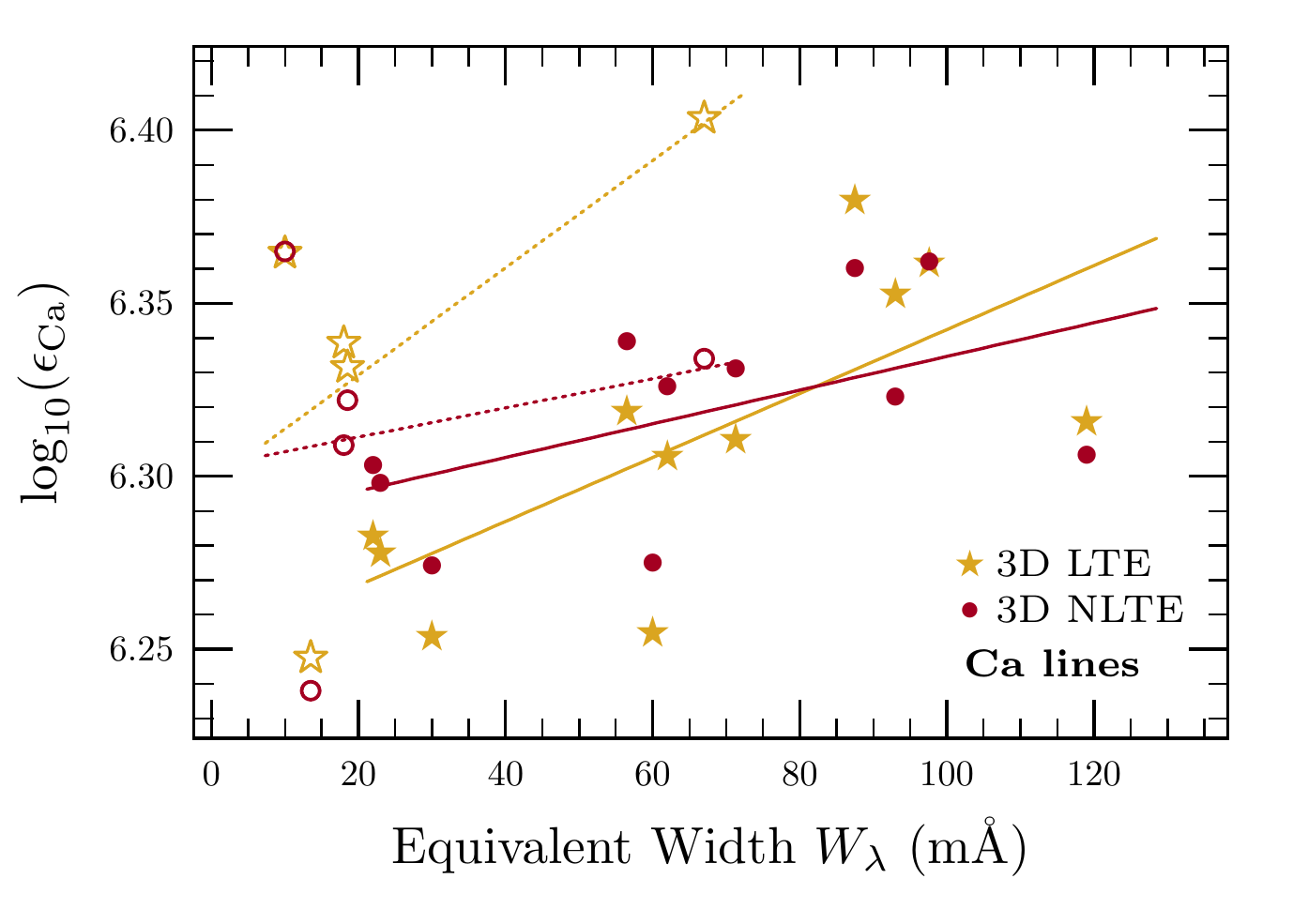}
\end{minipage}
\hspace{0.05\textwidth}
\begin{minipage}[t]{0.4\textwidth}
\centering
\includegraphics[width=\linewidth]{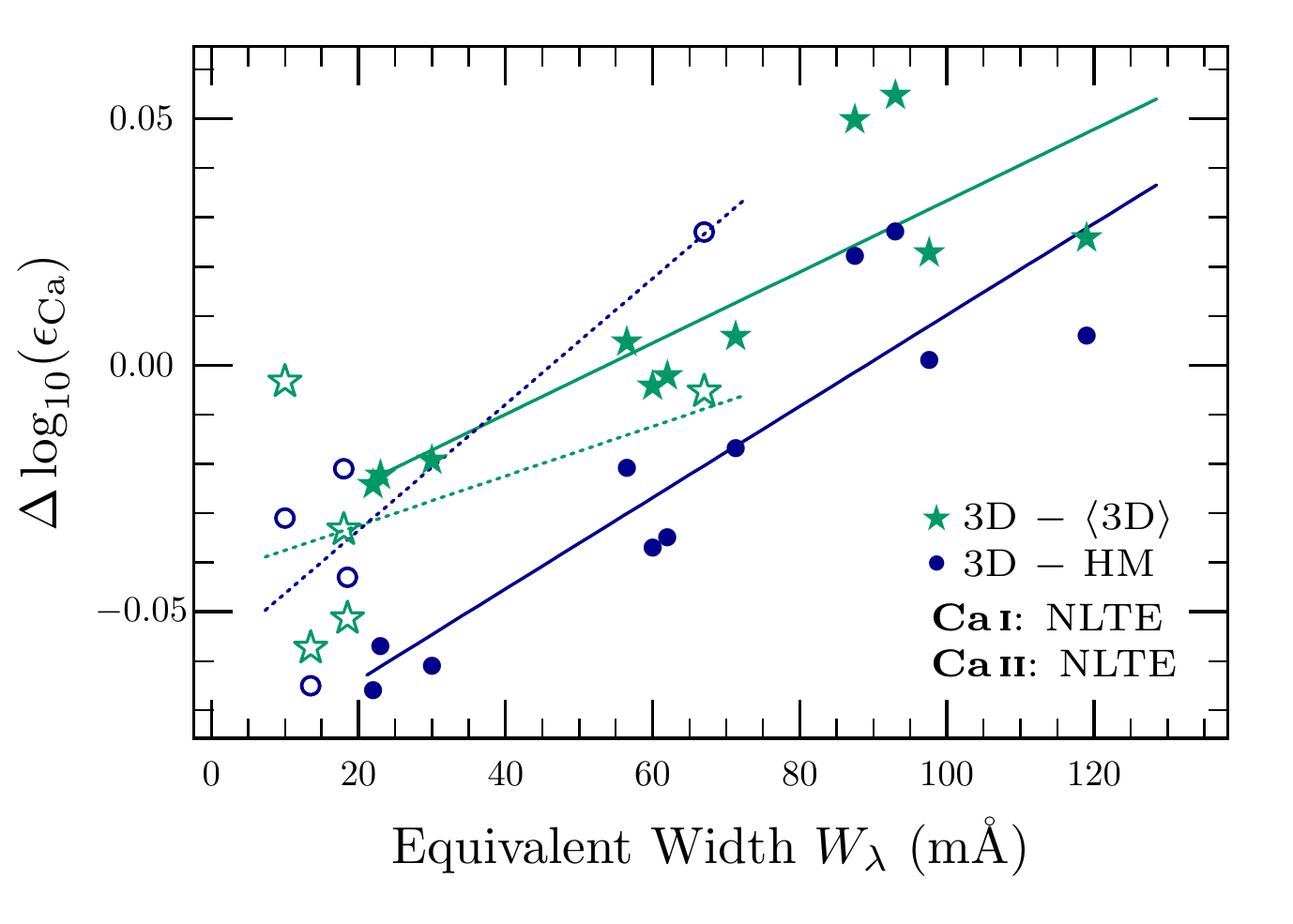}
\end{minipage}
\caption{\textit{Left}: Ca abundances derived from \cai and \caii lines with the 3D model, shown as a function of equivalent width.  \textit{Right}: Line-to-line differences between Ca abundances obtained with the 3D and \oneDAV\ models, and between those obtained with the 3D and HM models.  Filled symbols and solid trendlines indicate lines of the neutral species (\altcai), whereas open symbols and dotted lines indicate singly-ionised (\altcaii) lines.}
\label{fig:ca}
\end{figure*}

\subsection{Sulphur}
The line-to-line behaviour of our S abundances with line strength is shown in Fig.\ \ref{fig:s}.  The 3D abundance derived with the NLTE corrections given in Table~\ref{table:lines} is $\log \epsilon_{\mathrm{S}} = 7.13 \pm 0.03$ ($\pm$0.01 stat, $\pm$0.03 sys). The 1D models, with the exception of \marcs , return values very similar to the 3D analysis, consistent with the low sensitivity to the mean temperature structure and atmospheric inhomogeneities (Table~\ref{table:summary}).

Caffau et al.\ (\cite{caf2}) derived the solar abundance of sulphur with their own 3D model, using the IR triplet and two weaker lines that we also use (675.7 and 869.4\,nm). From equivalent measurements in the solar flux atlas, after discarding the discrepant 869.3\,nm line and taking into account departures from LTE estimated using a 1D model atmosphere, they found $\log\epsilon_{\mathrm{S}}= 7.25 \pm 0.08$ (s.d.).  It is not clear what caused the large differences in abundances between their two remaining weak lines and the IR triplet (0.19\,dex), as the somewhat smaller equivalent widths we measure in the flux atlas are not sufficient to explain these differences.  The problem cannot be attributed to erroneous $gf$-values either.

To investigate this further, we carried out a similar analysis using flux spectra for all of our \suli\ lines. We obtain agreement between the results from the weak and strong lines for slightly smaller $S_{\rm H}$ in flux than in intensity ($S_{\rm H}=0.1$ rather than $S_{\rm H}=0.3$). The mean abundance we infer for flux is $\log \epsilon_{\mathrm{S}} = 7.14 \pm 0.03$, in excellent agreement with the intensity results and between the two groups of lines. The reasons for the internal inconsistency of the results of Caffau et al.\ (\cite{caf2}) therefore remain unexplained.

The forbidden [\altsuli] 1082.1\,nm line has also sometimes been used as an abundance indicator. Because it is weak and originates from the ground level of \altsuli, it is formed in LTE. Caffau \& Ludwig (\cite{caf3}) measured equivalent widths of 0.22\,pm (disk-centre intensity) and 0.26\,pm (flux) for this line. With a very old $gf$-value taken from the NIST compilation, log $gf=-8.617$, they found a \suli abundance of $\log \epsilon_{\mathrm{S}} = 7.15$, in agreement with their results for the weak high-excitation lines, but much smaller than the IR triplet (Caffau et al.\ \cite{caf2}). We carefully rechecked the [\altsuli] line on various solar atlases. Our equivalent widths are always noticeably larger than the value of Caffau \& Ludwig (\cite{caf3}): $0.30 \pm 0.03$\,pm (intensity) and $0.36 \pm 0.03$\,pm (flux). Furthermore, a more recent and more accurate $gf$-value exists from C. Froese-Fischer: $\log gf= -8.774$\footnote{\href{http://atoms.vuse.vanderbilt.edu}{http://atoms.vuse.vanderbilt.edu} and private communication}. With our equivalent widths and the new $gf$-value, the inferred abundance of Caffau \& Ludwig becomes $\log\epsilon_{\mathrm{S}} = 7.44$, much larger than suggested by the permitted lines. We thus reject this forbidden line as most likely blended, even if we have not been able to identify a plausible perturbing line.

\setcounter{table}{4}

\begin{table*}[tbp]
\centering
\caption{A summary of the NLTE results obtained in this analysis with our 3D model, and with the four different 1D models we used.  The uncertainties given for the 3D results are the total errors, including statistical and systematic contributions, calculated as described in Sect.~\ref{s:atmospheres}.  For comparison, we also give the mean abundance difference between the 3D and 1D HM models, and between the 3D and $\langle 3{\rm D}\rangle$ (mean 3D) model.  Note that because all means were computed using abundances accurate to three decimal places, entries in columns 8 and 9 differ in some cases from the differences between the entries in columns 3--5.  We also give our final recommended abundance and the meteoritic values (Lodders et al.\ \cite{lodd}, normalised to the silicon abundance determined in Paper I).}
\label{table:summary}
\begin{tabular}{l l c c c c c c c c c}
\hline
\hline
& Species & 3D & \oneDAV & HM & \textsc{marcs} & \miss & 3D$-$HM & 3D$-$\oneDAV & Recommended & Meteoritic\\
\hline
$\log \epsilon_\mathrm{Na}$     & \nai  &  $6.21\pm0.04$ & 6.23 & 6.27 & 6.19 & 6.25 & $-$0.05 & $-$0.02 & $6.21\pm0.04$ & $6.27\pm0.02$ \vspace{1mm}\\
$\log \epsilon_\mathrm{Mg}$     & \mgi  &  $7.60\pm0.04$ & 7.61 & 7.64 & 7.56 & 7.63 & $-$0.04 & $-$0.01 &  \\
                                			& \mgii &  $7.58\pm0.05$ & 7.63 & 7.60 & 7.52 & 7.64 & $-$0.02 & $-$0.05 & \\
                                			& Mg all&  $7.59\pm0.04$ & 7.62 & 7.63 & 7.55 & 7.64 & $-$0.03 & $-$0.03 & $7.59\pm0.04$ & $7.53\pm0.01$ \vspace{1mm} \\
$\log \epsilon_\mathrm{Al}$     & \ali  &  $6.43\pm0.04$ & 6.45 & 6.48 & 6.40 & 6.47 & $-$0.05 & $-$0.02 & $6.43\pm0.04$ & $6.43\pm0.01$ \vspace{1mm}\\
$\log \epsilon_\mathrm{Si}$     & \sii  &  $7.51\pm0.03$ & 7.50 & 7.52 & 7.44 & 7.53 & $-$0.01 & $+$0.01 & \\
                                			& \siii &  $7.54\pm0.04$ & 7.58 & 7.55 & 7.47 & 7.60 & $-$0.02 & $-$0.04 & \\
                             			& Si all&  $7.51\pm0.03$ & 7.51 & 7.53 & 7.45 & 7.53 & $-$0.01 & \ph0.00 & $7.51\pm0.03$ & $7.51\pm0.01$ \vspace{1mm} \\
$\log \epsilon_\mathrm{P}$      & \phosi&  $5.41\pm0.03$ & 5.43 & 5.42 & 5.38 & 5.45 & $-$0.01 & $-$0.01 & $5.41\pm0.03$ & $5.43\pm0.04$ \vspace{1mm}\\
$\log \epsilon_\mathrm{S}$      & \suli &  $7.12\pm0.03$ & 7.14 & 7.12 & 7.06 & 7.15 & \ph0.00 & $-$0.02 & $7.12\pm0.03$ & $7.15\pm0.02$ \vspace{1mm}\\
$\log \epsilon_\mathrm{K}$      & \ki   &  $5.04\pm0.05$ & 5.06 & 5.11 & 5.02 & 5.09 & $-$0.06 & $-$0.02 & $5.04\pm0.05$ & $5.08\pm0.02$ \vspace{1mm}\\
$\log \epsilon_\mathrm{Ca}$     & \cai  &  $6.32\pm0.03$ & 6.31 & 6.34 & 6.26 & 6.30 & $-$0.02 & $+$0.01 & \\
                                			& \caii &  $6.32\pm0.04$ & 6.34 & 6.33 & 6.28 & 6.36 & $-$0.02 & $-$0.03 & \\
                                			& Ca all&  $6.32\pm0.03$ & 6.32 & 6.34 & 6.26 & 6.31 & $-$0.02 & \ph0.00 &  $6.32\pm0.03$ & $6.29\pm0.02$ \vspace{0.5mm}\\
\hline
\end{tabular}
\end{table*}

\subsection{Potassium}
Our mean solar K abundance is $\log\epsilon_{\mathrm{K}} = 5.04 \pm  0.07$ (s.d.), which with the full error budget described in Sect.~\ref{s:errors} becomes $\log\epsilon_{\mathrm{K}} = 5.04 \pm 0.05$ ($\pm 0.03$ stat, $\pm 0.04$ sys). We did not retain the very strong \ki line at 769.8974\,nm, because the abundance result from this line is very sensitive to uncertainties in the equivalent width and the large NLTE correction, which is of order 0.2\,dex. The result from this line is however consistent with the results from the other \ki lines.  Not surprisingly for these rather low-excitation lines, the HM model returns a significantly larger \ki abundance than in 3D, whereas the 3D-\oneDAV\ effects are minor.

Based on a 1D NLTE analysis, Zhang et al. (\cite{zhan}) derived a significantly larger solar abundance than ours: $\log\epsilon_{\mathrm{K}} = 5.12 \pm  0.03$ (s.d.).  The weak lines in common agree very well for the 1D theoretical model atmospheres.  The difference in mean abundance is mainly driven by their larger values for the IR lines, and by their inclusion of the strong \ki resonance line.  More recently, Caffau et al.\ (\cite{caf4}) presented a 3D LTE analysis of six \ki lines (including the strong 769.8\,nm resonance line), to which they added the 1D NLTE abundance corrections of Zhang et al. (\cite{zhan}). They obtained $\log\epsilon_{\mathrm{K}} = 5.11 \pm  0.10$ (s.d.), where the large scatter is conspicuous and largely driven by the 693.8\,nm line, for which they find an uncomfortably low abundance. We have not been able to reproduce their results for this line, even with their adopted equivalent width and $gf$-value.

\subsection{Calcium}
Line-to-line behaviour of our Ca abundances with equivalent width is shown in Fig.\ \ref{fig:ca}.  Our 3D+NLTE abundances of Ca are $\log\epsilon_{\mathrm{Ca}} = 6.32 \pm 0.03$ (s.d.; \altcai) and $\log\epsilon_{\mathrm{Ca}} = 6.32 \pm 0.04$ (s.d.; \altcaii). If we take all the \cai and \caii lines together, we have $\log\epsilon_{\mathrm{Ca}} = 6.32 \pm 0.03$ ($\pm$0.01 stat, $\pm$0.03 sys), with the uncertainty calculated as described in Sect.~\ref{s:errors}.  We note that the forbidden \caii line leads to an abundance consistent with the other lines ($\log\epsilon_{\mathrm{Ca}} = 6.365$).  The 3D and 1D results are very similar, with the exception again of the \marcs-based abundance, which is significantly lower (Table~\ref{table:summary}).  We have also measured the equivalent widths in the solar flux atlas, from which we have derived a solar Ca abundance in very good agreement with that for disk-centre intensity.  As explained above, intensity results are more definitive.

\subsection{Fluorine, Neon, Chlorine and Argon}
We do not attempt full redeterminations of the solar F, Ne, Cl and Ar abundances here. Ne and Ar are noble gases lacking spectral lines of photospheric origin in the solar spectrum, so only indirect solar abundance determinations can be attempted. For Ne and Ar we adopt the recommended abundances of AGSS09, to which we refer the interested reader for details on how these values are obtained.  Suitable lines of F and Cl are not present in the standard solar spectrum of the quiet Sun. Their abundances have however been estimated using IR HF and HCl lines in sunspot spectra.  The solar F abundance was recently rederived by Maiorca et al.\ (\cite{maiorca}; $\log\epsilon_{\rm F}=4.40\pm0.25$) using new sunspot observations, modern sunspot modelling and, crucially, laboratory molecular data (with clarifications following J\"onsson et al \cite{Jonsson14}).
The previous analysis (Hall \& Noyes \cite{hall_f}) relied on theoretical molecular data, and returned an abundance 0.16\,dex larger.  The new abundance agrees perfectly with the meteoritic value ($\log\epsilon_{\rm F}=4.42\pm0.06$, Lodders et al. \cite{lodd}; AGSS09).  A redetermination of the solar Cl abundance using similarly improved molecular data, observations and modelling has not yet been performed, with the current reference analysis now over 40 years old (Hall \& Noyes \cite{hall_cl}, see also AGSS09).

\section{Comparison with previous solar abundance compilations}
\label{s:discussion}

Table \ref{table:compilations} lists the recommended present-day solar photospheric abundances of the elements F to Ca, from some of the most commonly used compilations of the solar chemical composition. It is important to bear in mind that with the exception of AGSS09, all these sources are exactly that: inhomogeneous compilations of a multitude of literature sources using different model atmospheres, line formation techniques, atomic data, computer codes and error treatments. Although these studies have received a huge number of citations, they are in most part summaries of a huge amount of work done by atomic and molecular physicists providing the necessary input data, modellers of stellar atmospheres and spectral line formation, and solar physicists and spectroscopists carefully measuring the solar spectrum.  Without these inputs, none of those lauded solar abundance works (nor this series either) would have appeared. However, we draw particular attention to the fact that our work in AGSS09 is the only study in which a solar abundance analysis has been performed for all elements in a fully homogeneous manner. Furthermore, we have attempted to estimate the remaining systematic errors in detail, whereas almost all previous studies have accounted for the statistical errors only, either estimated through the standard deviation or the standard error, without consistency across different elements. 

As is apparent from  Table \ref{table:compilations}, the solar abundances we recommend here are very similar to those published in AGSS09, which is hardly surprising given that this paper is an update of the relevant part of that analysis. Here we have improved some of the elemental abundances following improved NLTE calculations and the appearance of better atomic data.  In some cases we have also slightly refined our line selections.  Apart from F, the maximum abundance difference is for Na ($-0.03$\,dex), Al and Ca ($-0.02$\,dex), in all cases well within the respective estimated uncertainties.  Compared with our preliminary analysis presented in AGS05, our new values are systematically higher by $\sim$$0.05$\,dex, although some elements behave differently (e.g. Ca). We attribute most of this difference to the steeper temperature gradient in the older 3D solar model atmosphere employed in AGS05.  However, it should be pointed out that for the elements Na to Ca, in AGS05 we relied only on published equivalent widths from Lambert and Luck (\cite{lamb1}) rather than measuring equivalent widths ourselves, as we have done here. This becomes particularly important for stronger, (partly) saturated lines, where it is paramount that the observed and the predicted equivalent widths are determined in a fully consistent manner. 

\begin{table}
\centering
\caption{Recommended present-day solar photospheric abundances for the intermediate-mass elements F to Ca, compared with oft-used solar abundance compilations: AG89 (Anders \& Grevesse \cite{ag89}), GS98 (Grevesse \& Sauval \cite{gs98}), AGS05 (Asplund et al. \cite{AGS05}), AGSS09 (Asplund et al. \cite{asp8}), LPG09 (Lodders et al. \cite{lodd}).  Preferred values are from this work except where noted.}
\label{table:compilations} 
\begin{tabular}{l@{\hspace{2mm}}l@{\hspace{3mm}}c@{\hspace{2mm}}c@{\hspace{3mm}}c@{\hspace{3mm}}c@{\hspace{2mm}}c@{\hspace{2mm}}c}
\hline \phantom{1}Z & el. & Preferred\phantom{\tablefootmark{1}} & AG89 & GS98 & AGS05 & AGSS09 & LPG09 \\
\hline
\hline
\phantom{1}9  &   F& $    4.40\pm 0.25 $\tablefootmark{\textcolor{BrickRed}{1}} &    4.56 &    4.56 &    4.56 &    4.56 &    4.56\\
10            &  Ne& $    7.93\pm 0.10 $\tablefootmark{\textcolor{BrickRed}{2}} &    8.09 &    8.08 &    7.84 &    7.93 &    8.05\\
11            &  Na& $    6.21\pm 0.04 $\phantom{\tablefootmark{1}}             &    6.33 &    6.33 &    6.17 &    6.24 &    6.30\\
12            &  Mg& $    7.59\pm 0.04 $\phantom{\tablefootmark{1}}             &    7.58 &    7.58 &    7.53 &    7.60 &    7.54\\
13            &  Al& $    6.43\pm 0.04 $\phantom{\tablefootmark{1}}             &    6.47 &    6.47 &    6.37 &    6.45 &    6.47\\
14            &  Si& $    7.51\pm 0.03 $\phantom{\tablefootmark{1}}             &    7.55 &    7.55 &    7.51 &    7.51 &    7.52\\
15            &   P& $    5.41\pm 0.03 $\phantom{\tablefootmark{1}}             &    5.45 &    5.45 &    5.36 &    5.41 &    5.46\\
16            &   S& $    7.12\pm 0.03 $\phantom{\tablefootmark{1}}             &    7.21 &    7.33 &    7.14 &    7.12 &    7.14\\
17            &  Cl& $    5.50\pm 0.30 $\tablefootmark{\textcolor{BrickRed}{3}} &    5.50 &    5.50 &    5.50 &    5.50 &    5.50\\
18            &  Ar& $    6.40\pm 0.13 $\tablefootmark{\textcolor{BrickRed}{2}} &    6.56 &    6.40 &    6.18 &    6.40 &    6.50\\
19            &   K& $    5.04\pm 0.05 $\phantom{\tablefootmark{1}}             &    5.12 &    5.12 &    5.08 &    5.03 &    5.12\\
20            &  Ca& $    6.32\pm 0.03 $\phantom{\tablefootmark{1}}             &    6.36 &    6.36 &    6.31 &    6.34 &    6.33\\
\hline
\end{tabular}
\tablefoot{\tablefoottext{1}{Maiorca et al.\ (\cite{maiorca})}; \tablefoottext{2}{AGSS09}; \tablefoottext{3}{Hall \& Noyes (\cite{hall_cl})}}
\end{table}

Most solar abundances for these intermediate-mass elements have remained rather steady since the publication of the work on the solar system chemical composition by Anders \& Grevesse (\cite{ag89}).  Some elements have seen substantial differences though, most notably F ($-0.16$\,dex), Na ($-0.12$\,dex) and the noble gases Ne and Ar, which rely on the rather dramatic negative revision of the solar oxygen abundance in recent years (e.g. Allende Prieto et al. \cite{cap_o}; Asplund et al. \cite{asp4}; Pereira et al. \cite{pereira_o}; AGSS09). Sulphur and potassium have also decreased significantly ($-0.08$\,dex) relative to Anders \& Grevesse (\cite{ag89}), and several other elements have decreased slightly ($\approx$$-0.04$\,dex).  This includes Si, which necessitates an overall shift in all meteoritic abundances, as they are all measured relative to Si (e.g. Anders \& Grevesse \cite{ag89}; Asplund \cite{asp3}; AGSS09).  We note that the oft-used solar abundance compilation of Grevesse \& Sauval (\cite{gs98}) is largely the same as Anders \& Grevesse (\cite{ag89}), with the main exceptions being the crucial elements C, N, O and Fe, following an empirical and rather \textit{ad hoc} adjustment to the HM temperature structure to remove Fe abundance trends with excitation potential.

Lodders et al. (\cite{lodd}) carried out a meticulous evaluation of the meteoritic abundances, but also presented photospheric abundances taken from the literature. For the intermediate elements, they systematically avoided selecting any of the recommended values from AGS05, as they argued that those abundances were too low compared to the meteoritic values on their favoured absolute scale, and that the 3D solar model was not sufficiently tested. However, they did choose to adopt the P and S abundances from the 3D analyses of Caffau et al. (\cite{caf1}, \cite{caf2}), as well as several analyses based on 1D theoretical model atmospheres that are known not to reproduce the solar temperature structure very well. As we will discuss in detail in a subsequent paper in this series, selecting photospheric results partly on the similarity with the meteoritic evidence is not advisable, and may easily prejudice findings. As shown by Pereira et al.\ (\cite{pereira_o}), the 3D solar model employed in AGS05 in fact already equalled the performance of the HM model and outperformed all theoretical 1D models in all observational tests -- and the improved version we use here does even better (Pereira et al.\ \cite{pereira_models}).

\section{Conclusions}
\label{s:conclusion}

We have presented a comprehensive determination of the solar elemental abundances of the intermediate-mass elements Na to Ca, scrutinising all ingredients in the analysis to a high degree. We have carefully scoured the literature for the best possible atomic data to use (line identifications, excitation potentials, transition probabilities, isotopic shifts, HFS, ionisation energies, partition functions, etc.).  We have been extremely selective in choosing which spectral lines to employ, discarding all candidates that showed \textit{a priori} deficiencies: too weak, too strong, known or suspected to be blended, or only usable with questionable atomic data. In particular, we preferred to err on the side of caution regarding blends, choosing to exclude any lines deviating from the well-known unblended ``C'' bisector shape (indicating undetected blending), or when the equivalent width was impossible to measure with high accuracy.  Including dubious lines almost always skews results upwards due to blends, making lines appear stronger and increasing the line-to-line scatter.  We gave all lines an individual weighting based on their observed profiles and available atomic data.  The low abundance scatter we find points to the soundness of this strategy.

We analysed all lines using a highly realistic, time-dependent, 3D, hydrodynamic model atmosphere of the solar photosphere and subsurface convection that treats the crucial radiative heating and cooling in detail using the best available continuous and line opacities.  As shown by Pereira et al.\ (\cite{pereira_models}), our 3D solar model has successfully passed a multitude of observational tests, demonstrating that is superior to any 1D model, theoretical or semi-empirical, and at least as good as other realistic, state-of-the-art 3D models (Beeck et al. \cite{beeck2012}). In particular, our \textit{ab initio} 3D model outperforms the HM model in the continuum centre-to-limb variation, even though the temperature structure of the HM model was initially constructed to reproduce that key diagnostic. It is clear from Table~\ref{table:summary} that the sensitivity of the solar photospheric abundances of Na to Ca to the actual model atmosphere is however rather modest: with the exception of the 1D theoretical \marcs\ model, all 1D models return Na-Ca abundances typically within $0.05$\,dex of the 3D values.  The venerable and traditionally-used HM model always yields abundances higher than those recommended here. It is reassuring that the \oneDAV-based results are very similar to the full 3D case for these elements ($\le$$0.03$\,dex difference).  The full 3D effect for the Sun for these elements is thus primarily the result of a different mean atmospheric stratification to previous models, rather than the presence of atmospheric inhomogeneities.  This opens the way for accurate stellar abundance determinations of these elements to be obtained in a straightforward manner with spatially- and temporally-averaged 3D models and existing 1D spectrum synthesis machinery, at least in solar-type stars. Such \oneDAV\ spectrum modelling has started to be used in the analyses of observed stellar spectra (e.g. Bergemann et al. \cite{Bergemann2012}).  Further verification is needed before the same can be said for stars with significantly different effective temperatures, metallicities or surface gravities to the Sun, however.

We have accounted for NLTE effects in the line formation whenever possible. Full 3D NLTE calculations are still only available for a few elements such as Li, O, Na and Ca, for a very small number of stars (e.g. Asplund et al. \cite{asp4}; Pereira et al. \cite{pereira_o}; Lind et al. \cite{lind_li6}).  Work is however in progress to extend this to additional elements and different stellar parameters. As a substitute, we have included NLTE abundance corrections computed with various 1D model atmospheres, in most cases carried out by ourselves but in a few cases taken from the literature. For our recommended 3D+NLTE solar abundances we have considered the departures from LTE for the spatially and temporally-averaged 3D model \oneDAV, which should show similar behaviour to the full 3D model in this respect, due to the fact that radiative transfer primarily takes place vertically (Lind et al. \cite{lind_li6}).  From Table~\ref{table:lines} we see that the NLTE corrections are in any case rather small for our selected solar lines (with a few notable exceptions). We stress however that the NLTE results are a significant improvement over the pure 3D LTE results, as in many cases they remove abundance trends and decrease the abundance scatter.  

We have also taken HFS into account for \altnai, \ali and \altki. The effect of this additional broadening turns out to be very small, resulting in final changes of $<$$0.01$\,dex in the derived abundances. The main reason for this is that the lines we use from these elements are typically weak. However, the HFS does undoubtedly play a role in the shapes of the line profiles.

Finally, we have spent considerable effort quantifying not only the statistical uncertainties, but more importantly, the remaining systematic errors due to possible shortcomings in the atmospheric modelling and NLTE line formation. When relevant we have also discussed in detail the possible impact of uncertainties in the atomic data and how that may influence our recommended solar photospheric abundances. This series of articles is the first time that detailed abundances of all elements have been determined homogeneously with a 3D-based analysis, and with a proper quantitative treatment of the abundance uncertainties.

Our recommended present-day solar photospheric abundances and associated uncertainties for Na to Ca are summarised in Table \ref{table:summary}.  We are confident that these represent a significant improvement over previous determinations, and are thus the best possible values to be used by the astronomical community at the present time.

\begin{acknowledgements}
We thank Katia Cunha, Enrico Maiorca and Nils Ryde for helpful discussions concerning the transition probabilities and partition functions of HF, Thomas Gehren for providing unpublished NLTE abundance corrections, Charlotte Froese-Fischer for other helpful discussions on transition probabilities, and the referee for constructive feedback. PS, NG and MA variously thank the Max Planck Institut f\"ur Astrophysik, Garching, the Centre Spatial de Li\`ege, the Department of Astrophysics, Geophysics and Oceanography, University of Li\`ege and Mount Stromlo Observatory for support and hospitality during the production of this paper.  We acknowledge further support from IAU Commission 46, the Lorne Trottier Chair in Astrophysics, the (Canadian) Institute for Particle Physics, the Banting Fellowship scheme as administered by the Natural Science and Engineering Research Council of Canada, the UK Science \& Technology Facilities Council (PS), the Royal Belgian Observatory (NG), the Australian Research Council (MA) and the Australian Research Council's DECRA scheme (project number DE120102940; RC).
\end{acknowledgements}

\setcounter{table}{0}

\Online
\onecolumn

\setcounter{table}{0}

\begin{table*}[htbp]
\centering
\caption{The mean stratification and rms variations of the 3D hydrodynamic model atmosphere that we employ here. All quantities are averaged over surfaces of common $\tau_{500\,{\rm nm}}$.}
\label{table:3Dmodel}
\begin{tabular}{l c c c c c c c c}
\hline
\hline
$\log\tau_{500\,{\rm nm}}$ & $T$ & $\Delta T_{\rm rms}$ & $\rho$ & $\Delta \rho_{\rm rms}$ &
$P_{\rm gas}$ & $\Delta P_{\rm gas, rms}$ &
$v_{\rm z}$ & $\Delta v_{\rm z,rms}$ \\
& (K) & (K) & (kg/m$^3$) & (kg/m$^3$) & (Pa)   & (Pa)  & (km/s) & (km/s) \\
\hline
 $-5.00$ &    4141 &    449 &   8.61E-07 &   1.21E-07 &   2.35E+01 &   2.95E+00  &     0.08 &     1.44 \\
 $-4.80$ &    4165 &    444 &   1.12E-06 &   1.60E-07 &   3.07E+01 &   3.68E+00  &     0.07 &     1.37 \\
 $-4.60$ &    4194 &    435 &   1.45E-06 &   2.13E-07 &   4.02E+01 &   4.77E+00  &     0.07 &     1.29 \\
 $-4.40$ &    4226 &    425 &   1.89E-06 &   2.78E-07 &   5.27E+01 &   6.17E+00  &     0.06 &     1.22 \\
 $-4.20$ &    4260 &    412 &   2.46E-06 &   3.56E-07 &   6.90E+01 &   7.86E+00  &     0.05 &     1.14 \\
 $-4.00$ &    4296 &    398 &   3.19E-06 &   4.49E-07 &   9.03E+01 &   9.85E+00  &     0.05 &     1.07 \\
 $-3.80$ &    4333 &    383 &   4.14E-06 &   5.60E-07 &   1.18E+02 &   1.22E+01  &     0.05 &     1.01 \\
 $-3.60$ &    4370 &    368 &   5.34E-06 &   6.89E-07 &   1.54E+02 &   1.48E+01  &     0.05 &     0.95 \\
 $-3.40$ &    4407 &    351 &   6.89E-06 &   8.38E-07 &   2.00E+02 &   1.79E+01  &     0.05 &     0.90 \\
 $-3.20$ &    4444 &    331 &   8.87E-06 &   1.00E-06 &   2.60E+02 &   2.14E+01  &     0.05 &     0.86 \\
 $-3.00$ &    4483 &    311 &   1.14E-05 &   1.19E-06 &   3.37E+02 &   2.54E+01  &     0.05 &     0.83 \\
 $-2.80$ &    4524 &    291 &   1.46E-05 &   1.40E-06 &   4.36E+02 &   2.99E+01  &     0.06 &     0.80 \\
 $-2.60$ &    4569 &    275 &   1.87E-05 &   1.62E-06 &   5.64E+02 &   3.50E+01  &     0.06 &     0.80 \\
 $-2.40$ &    4618 &    262 &   2.39E-05 &   1.87E-06 &   7.28E+02 &   4.07E+01  &     0.07 &     0.80 \\
 $-2.20$ &    4670 &    251 &   3.04E-05 &   2.15E-06 &   9.39E+02 &   4.73E+01  &     0.07 &     0.82 \\
 $-2.00$ &    4724 &    239 &   3.87E-05 &   2.46E-06 &   1.21E+03 &   5.54E+01  &     0.08 &     0.85 \\
 $-1.80$ &    4783 &    224 &   4.92E-05 &   2.82E-06 &   1.55E+03 &   6.63E+01  &     0.09 &     0.91 \\
 $-1.60$ &    4849 &    206 &   6.23E-05 &   3.26E-06 &   2.00E+03 &   8.16E+01  &     0.09 &     0.98 \\
 $-1.40$ &    4927 &    185 &   7.86E-05 &   3.86E-06 &   2.56E+03 &   1.04E+02  &     0.10 &     1.07 \\
 $-1.20$ &    5023 &    162 &   9.88E-05 &   4.77E-06 &   3.28E+03 &   1.37E+02  &     0.10 &     1.19 \\
 $-1.00$ &    5144 &    138 &   1.23E-04 &   6.20E-06 &   4.19E+03 &   1.88E+02  &     0.10 &     1.33 \\
 $-0.80$ &    5299 &    118 &   1.52E-04 &   8.35E-06 &   5.34E+03 &   2.65E+02  &     0.09 &     1.49 \\
 $-0.60$ &    5496 &    120 &   1.85E-04 &   1.15E-05 &   6.73E+03 &   3.76E+02  &     0.08 &     1.67 \\
 $-0.40$ &    5740 &    159 &   2.19E-04 &   1.65E-05 &   8.31E+03 &   5.30E+02  &     0.07 &     1.87 \\
 $-0.20$ &    6044 &    230 &   2.50E-04 &   2.45E-05 &   9.96E+03 &   7.59E+02  &     0.05 &     2.08 \\
 $+0.00$ &    6422 &    322 &   2.73E-04 &   3.56E-05 &   1.15E+04 &   1.10E+03  &     0.01 &     2.30 \\
   +0.20 &    6884 &    441 &   2.85E-04 &   4.89E-05 &   1.29E+04 &   1.55E+03  &    -0.05 &     2.52 \\
   +0.40 &    7430 &    589 &   2.89E-04 &   6.29E-05 &   1.40E+04 &   2.08E+03  &    -0.16 &     2.73 \\
   +0.60 &    8021 &    729 &   2.86E-04 &   7.57E-05 &   1.49E+04 &   2.64E+03  &    -0.29 &     2.92 \\
   +0.80 &    8587 &    823 &   2.81E-04 &   8.59E-05 &   1.58E+04 &   3.20E+03  &    -0.40 &     3.06 \\
   +1.00 &    9080 &    856 &   2.77E-04 &   9.35E-05 &   1.66E+04 &   3.73E+03  &    -0.49 &     3.15 \\
   +1.20 &    9479 &    832 &   2.76E-04 &   9.91E-05 &   1.74E+04 &   4.27E+03  &    -0.55 &     3.19 \\
   +1.40 &    9790 &    767 &   2.78E-04 &   1.03E-04 &   1.84E+04 &   4.80E+03  &    -0.59 &     3.19 \\
   +1.60 &   10037 &    683 &   2.84E-04 &   1.06E-04 &   1.94E+04 &   5.33E+03  &    -0.60 &     3.16 \\
   +1.80 &   10247 &    598 &   2.94E-04 &   1.09E-04 &   2.07E+04 &   5.86E+03  &    -0.60 &     3.10 \\
   +2.00 &   10442 &    522 &   3.07E-04 &   1.11E-04 &   2.23E+04 &   6.36E+03  &    -0.59 &     3.03 \\
\hline
\end{tabular}
\end{table*}

\clearpage

\begin{center}
\topcaption{\label{table:lines} Lines retained in this analysis: atomic and solar data, line weightings, LTE abundance results for the five models used in this analysis, NLTE corrections to the LTE result (when available), and the corresponding 3D+NLTE abundance result}
\tablefirsthead{%
  \hline
  $\lambda$ & & E$_{exc}$ & $\log$ $gf$ & $gf$ & $W_\lambda$ & Wt.  &\multicolumn{5}{c}{LTE Abundances}& $\Delta_{\rm NLTE}$ & 3D \\
  (nm) & & (eV) & & ref. & (pm) &  & 3D & $\langle 3{\rm D}\rangle$ & HM & \marcs & \miss & (3D) & NLTE\\
  \hline
   & & & & & & & & & & & & & \\
  }
\tablehead{%
  \hline 
  $\lambda$ & & E$_{exc}$ & $\log$ $gf$ & $gf$ & $W_\lambda$ & Wt.  &\multicolumn{5}{c}{LTE Abundances}& $\Delta_{\rm NLTE}$ & 3D \\
  (nm) & & (eV) & & ref. & (pm) &  & 3D & $\langle 3{\rm D}\rangle$ & HM & \marcs & \miss & (3D) & NLTE\\
  \hline
  \multicolumn{14}{r}{continued.}\\
  \hline
   & & & & & & & & & & & & & \\
  }
\tabletail{%
  \hline
  \multicolumn{14}{r}{continued on next page}\\
  \hline
}
\tablelasttail{\hline}
\begin{mpsupertabular}{r@{}c c c c r c c c c c c r c}
\multicolumn{14}{c} {\nai} \\
  475.1822 & & 2.104 & $-2.078$ & 1 & 1.110 & 1 & 6.232 & 6.254 & 6.296 & 6.213 & 6.282 & $-0.02$ & 6.212 \\
  514.8838 & & 2.102 & $-2.044$ & 1 & 1.270 & 1 & 6.243 & 6.265 & 6.307 & 6.221 & 6.293 & $-0.04$ & 6.203 \\
  615.4225 & & 2.102 & $-1.547$ & 1 & 3.730 & 1 & 6.275 & 6.295 & 6.336 & 6.243 & 6.318 & $-0.05$ & 6.225 \\
  616.0747 & & 2.104 & $-1.246$ & 1 & 5.700 & 1 & 6.264 & 6.280 & 6.321 & 6.221 & 6.300 & $-0.06$ & 6.204 \\
 1074.6440 & & 3.191 & $-1.294$ & 1 & 1.330 & 1 & 6.223 & 6.241 & 6.268 & 6.192 & 6.264 & $ 0.00$ & 6.223 \\
   & & & & & & & & & & & & & \\

\multicolumn{14}{c} {\mgi} \\
  631.8717 & & 5.108 & $-2.103$ & 2 & 4.130 & 1 & 7.665 & 7.676 & 7.707 & 7.620 & 7.698 & $+0.01$ & 7.675  \\
  631.9237 & & 5.108 & $-2.324$ & 2 & 2.600 & 1 & 7.622 & 7.634 & 7.667 & 7.583 & 7.658 & $+0.01$ & 7.632  \\
  892.3570 & & 5.394 & $-1.678$ & 2 & 6.330 & 2 & 7.651 & 7.658 & 7.685 & 7.596 & 7.673 & $+0.01$ & 7.661  \\
  942.9810 & & 5.932 & $-1.271$ & 2 & 4.710 & 1 & 7.471 & 7.482 & 7.504 & 7.427 & 7.500 & $ 0.00$ & 7.471  \\
  998.3190 & & 5.932 & $-2.153$ & 2 & 1.000 & 2 & 7.541 & 7.554 & 7.581 & 7.505 & 7.576 & $ 0.00$ & 7.541  \\
 1031.2520 & & 6.118 & $-1.730$ & 3 & 1.830 & 1 & 7.549 & 7.561 & 7.586 & 7.511 & 7.584 & $ 0.00$ & 7.549  \\
 1152.2210 & & 6.118 & $-1.910$ & 3 & 2.100 & 1 & 7.702 & 7.714 & 7.737 & 7.667 & 7.736 & $ 0.00$ & 7.702  \\
 1241.7910 & & 5.932 & $-1.664$ & 2 & 4.480 & 2 & 7.592 & 7.606 & 7.634 & 7.551 & 7.630 & $ 0.00$ & 7.592  \\
 1242.3000 & & 5.932 & $-1.188$ & 2 & 9.700 & 1 & 7.599 & 7.609 & 7.635 & 7.540 & 7.629 & $ 0.00$ & 7.599  \\
 1587.9521 & \footnote{Triplet computed as a single line; also contains components at 1587.9567\,nm ($\log gf=-1.250$) and 1587.9599\,nm ($\log gf=-3.175$)}
             & 5.946 & $-1.998$ & 2 &16.800 & 1 & 7.578 & 7.607 & 7.644 & 7.540 & 7.634 & $ 0.00$ & 7.578  \\
 1588.6183 & \footnote{Doublet computed as a single line; also contains component at 1588.6261\,nm ($\log gf=-1.396$)}
             & 5.946 & $-1.521$ & 2 &12.000 & 1 & 7.542 & 7.573 & 7.611 & 7.518 & 7.602 & $ 0.00$ & 7.542  \\
   & & & & & & & & & & & & & \\

\multicolumn{14}{c} {\mgii} \\
  787.7050 & & 9.996 & $ 0.391$ & 4 & 1.900 & 1 & 7.615 & 7.678 & 7.657 & 7.565 & 7.692 & $-0.03$ & 7.585  \\
  789.6400 & & 9.999 & $ 0.643$ & 4 & 3.000 & 1 & 7.707 & 7.758 & 7.730 & 7.642 & 7.769 & $-0.04$ & 7.667  \\
  921.8250 & & 8.655 & $ 0.268$ & 4 & 7.400 & 1 & 7.704 & 7.720 & 7.675 & 7.609 & 7.726 & $-0.08$ & 7.624  \\
  924.4270 & & 8.655 & $-0.034$ & 4 & 5.150 & 1 & 7.635 & 7.662 & 7.624 & 7.556 & 7.671 & $-0.06$ & 7.575  \\
 1009.2095 & \footnote{Triplet computed as a single line; also contains two components at 1009.2217\,nm ($\log gf=+1.020$ and $-0.530$)}
             &11.630 & $ 0.910$ & 5 & 1.330 & 1 & 7.559 & 7.677 & 7.656 & 7.542 & 7.693 & $ 0.00$ & 7.559  \\
 1091.4230 & & 8.864 & $ 0.038$ & 4 & 5.220 & 2 & 7.572 & 7.598 & 7.568 & 7.492 & 7.610 & $-0.05$ & 7.522  \\
   & & & & & & & & & & & & & \\

\multicolumn{14}{c} {\ali} \\
 669.6023 & & 3.143 & $-1.569$ & 6 & 3.580 & 2 & 6.435 & 6.453 & 6.490 & 6.400 & 6.474 & $+0.03$ & 6.465 \\
 669.8673 & & 3.143 & $-1.870$ & 6 & 2.100 & 3 & 6.441 & 6.460 & 6.498 & 6.412 & 6.483 & $+0.01$ & 6.451 \\
 783.5309 & & 4.022 & $-0.689$ & 6 & 4.100 & 1 & 6.372 & 6.385 & 6.412 & 6.330 & 6.405 & $ 0.00$ & 6.372 \\
 891.2900 & & 4.085 & $-1.963$ & 6 & 0.300 & 1 & 6.385 & 6.402 & 6.429 & 6.355 & 6.425 & $ 0.00$ & 6.385 \\
1076.8365 & & 4.085 & $-2.020$ & 6 & 0.400 & 1 & 6.452 & 6.469 & 6.497 & 6.423 & 6.492 & $ 0.00$ & 6.452 \\
1087.2973 & & 4.085 & $-1.326$ & 6 & 1.530 & 1 & 6.359 & 6.375 & 6.403 & 6.327 & 6.398 & $+0.02$ & 6.379 \\
1089.1736 & & 4.087 & $-1.027$ & 6 & 3.160 & 1 & 6.427 & 6.443 & 6.470 & 6.389 & 6.464 & $ 0.00$ & 6.427 \\
   & & & & & & & & & & & & & \\

\multicolumn{14}{c} {\sii} \\
 564.5613 & & 4.930 & $-2.043$ & 7 & 3.500 & 1 & 7.507 & 7.503 & 7.528 & 7.451 & 7.530 & $-0.01$ & 7.497  \\
 568.4484 & & 4.954 & $-1.553$ & 7 & 6.370 & 2 & 7.464 & 7.452 & 7.476 & 7.388 & 7.475 & $-0.04$ & 7.424  \\
 569.0425 & & 4.930 & $-1.773$ & 7 & 5.260 & 3 & 7.510 & 7.502 & 7.526 & 7.442 & 7.527 & $-0.01$ & 7.500  \\
 570.1105 & & 4.930 & $-1.953$ & 7 & 3.950 & 3 & 7.491 & 7.486 & 7.511 & 7.433 & 7.513 & $-0.01$ & 7.481  \\
 577.2145 & & 5.082 & $-1.653$ & 7 & 5.600 & 2 & 7.563 & 7.555 & 7.577 & 7.493 & 7.579 & $ 0.00$ & 7.563  \\
 579.3073 & & 4.930 & $-1.963$ & 7 & 4.580 & 1 & 7.601 & 7.594 & 7.618 & 7.538 & 7.619 & $-0.01$ & 7.591  \\
 674.1640 & & 5.984 & $-1.653$ & 7 & 1.630 & 1 & 7.614 & 7.615 & 7.631 & 7.569 & 7.639 & $-0.01$ & 7.604  \\
 703.4903 & & 5.871 & $-0.783$ & 7 & 7.400 & 1 & 7.552 & 7.541 & 7.552 & 7.470 & 7.558 & $-0.02$ & 7.532  \\
 722.6206 & & 5.614 & $-1.413$ & 7 & 3.870 & 1 & 7.508 & 7.500 & 7.515 & 7.444 & 7.522 & $-0.01$ & 7.498  \\
   & & & & & & & & & & & & & \\

\multicolumn{14}{c} {\siii} \\
 637.1370 & & 8.121 & $-0.044$ & 8 & 3.660 & 1 & 7.539 & 7.577 & 7.555 & 7.473 & 7.597 & $ 0.00$ & 7.539 \\
   & & & & & & & & & & & & & \\

\multicolumn{14}{c} {\phosi} \\
  952.5741 & & 6.985 & $-0.100$ & 9 & 0.700 & 1 & 5.422 & 5.437 & 5.432 & 5.393 & 5.456 &        &   \\
  975.0748 & & 6.954 & $-0.180$ & 9 & 0.660 & 1 & 5.433 & 5.448 & 5.443 & 5.404 & 5.467 &        &   \\
 1051.1588 & & 6.936 & $-0.130$ & 9 & 0.810 & 1 & 5.410 & 5.423 & 5.419 & 5.380 & 5.443 &        &   \\
 1052.9524 & & 6.954 & $+0.240$ & 9 & 1.570 & 1 & 5.393 & 5.404 & 5.397 & 5.355 & 5.424 &        &   \\
 1058.1577 & & 6.985 & $+0.450$ & 9 & 2.400 & 1 & 5.443 & 5.452 & 5.443 & 5.397 & 5.471 &        &   \\
 1059.6903 & & 6.936 & $-0.210$ & 9 & 0.750 & 1 & 5.446 & 5.460 & 5.457 & 5.418 & 5.479 &        &   \\
 1068.1406 & & 6.954 & $-0.190$ & 9 & 0.700 & 1 & 5.400 & 5.414 & 5.410 & 5.371 & 5.434 &        &   \\
 1081.3146 & & 6.985 & $-0.410$ & 9 & 0.400 & 1 & 5.369 & 5.384 & 5.382 & 5.344 & 5.405 &        &   \\
   & & & & & & & & & & & & & \\

\multicolumn{14}{c} {\suli} \\
  469.4113 & & 6.525 & $-1.673$ &10 & 1.210 & 1 & 7.177 & 7.211 & 7.221 & 7.170 & 7.236 & $-0.01$ & 7.169\\
  469.5443 & & 6.525 & $-1.829$ &10 & 0.870 & 1 & 7.164 & 7.198 & 7.208 & 7.161 & 7.222 & $-0.01$ & 7.156\\
  675.7171 & \footnote{Triplet computed as a single line; also contains components at 675.6851\,nm ($\log gf=-1.784$) and 675.7007\,nm ($\log gf=-0.934$)}
             & 7.870 & $-0.353$ &11 & 2.330 & 1 & 7.205 & 7.243 & 7.234 & 7.177 & 7.262 & $-0.01$ & 7.195\\
  867.0183 & & 7.866 & $-0.879$ &10 & 0.600 & 2 & 7.114 & 7.158 & 7.149 & 7.096 & 7.175 & $-0.01$ & 7.101\\
  869.4626 & & 7.870 & $+0.101$ &10 & 3.400 & 3 & 7.114 & 7.144 & 7.126 & 7.062 & 7.158 & $-0.02$ & 7.095\\
 1045.5449 & & 6.860 & $+0.250$ &12 &13.400 & 2 & 7.223 & 7.210 & 7.185 & 7.105 & 7.201 & $-0.10$ & 7.121\\
 1045.6757 & & 6.860 & $-0.447$ &12 & 6.200 & 1 & 7.157 & 7.158 & 7.134 & 7.080 & 7.172 & $-0.05$ & 7.108\\
 1045.9406 & & 6.860 & $+0.030$ &12 &10.600 & 2 & 7.194 & 7.183 & 7.157 & 7.095 & 7.191 & $-0.09$ & 7.109\\
   & & & & & & & & & & & & & \\

\multicolumn{14}{c} {\ki} \\
 404.4142 & & 0.000 & $-1.944$ &13 & 1.230 & 1 & 5.002 & 5.031 & 5.089 & 5.003 & 5.057 & $-0.04$ & 4.962 \\
 580.1749 & & 1.617 & $-1.605$ &14 & 0.175 & 1 & 5.173 & 5.194 & 5.229 & 5.149 & 5.220 & $-0.03$ & 5.143 \\
 693.8763 & & 1.617 & $-1.250$ &15 & 0.380 & 1 & 5.110 & 5.131 & 5.166 & 5.087 & 5.156 & $-0.03$ & 5.080 \\
1176.9639 & & 1.617 & $-0.452$ &14 & 3.300 & 1 & 5.056 & 5.074 & 5.113 & 5.026 & 5.096 & $-0.06$ & 4.996 \\
1252.2134 & & 1.617 & $-0.150$ &14 & 7.000 & 1 & 5.111 & 5.130 & 5.169 & 5.071 & 5.150 & $-0.08$ & 5.031 \\
   & & & & & & & & & & & & & \\

\multicolumn{14}{c} {\cai} \\
 451.2268 & & 2.526 & $-1.901$ &16 & 2.200 & 2 & 6.283 & 6.307 & 6.359 & 6.266 & 6.335 & $+0.02$ & 6.303 \\
 526.0387 & & 2.521 & $-1.719$ &17 & 3.000 & 2 & 6.254 & 6.273 & 6.325 & 6.225 & 6.299 & $+0.02$ & 6.274 \\
 586.7562 & & 2.933 & $-1.570$ &18 & 2.300 & 2 & 6.278 & 6.300 & 6.345 & 6.252 & 6.325 & $+0.02$ & 6.298 \\
 616.1297 & & 2.523 & $-1.266$ &17 & 6.000 & 1 & 6.255 & 6.259 & 6.312 & 6.200 & 6.274 & $+0.02$ & 6.275 \\
 616.3755 & & 2.521 & $-1.286$ &17 & 6.200 & 2 & 6.306 & 6.308 & 6.361 & 6.248 & 6.323 & $+0.02$ & 6.326 \\
 616.6439 & & 2.521 & $-1.142$ &17 & 7.130 & 3 & 6.311 & 6.305 & 6.358 & 6.241 & 6.316 & $+0.02$ & 6.331 \\
 616.9042 & & 2.523 & $-0.797$ &17 & 9.760 & 2 & 6.362 & 6.339 & 6.391 & 6.265 & 6.345 & $ 0.00$ & 6.362 \\
 616.9563 & & 2.526 & $-0.478$ &17 &11.900 & 2 & 6.316 & 6.290 & 6.340 & 6.210 & 6.296 & $-0.01$ & 6.306 \\
 645.5598 & & 2.523 & $-1.340$ &18 & 5.650 & 2 & 6.319 & 6.314 & 6.370 & 6.258 & 6.327 & $+0.02$ & 6.339 \\
 647.1662 & & 2.526 & $-0.686$ &17 & 9.300 & 3 & 6.353 & 6.298 & 6.356 & 6.230 & 6.297 & $-0.03$ & 6.323 \\
 649.9650 & & 2.523 & $-0.818$ &17 & 8.750 & 3 & 6.380 & 6.330 & 6.388 & 6.264 & 6.330 & $-0.02$ & 6.360 \\
   & & & & & & & & & & & & & \\

\multicolumn{14}{c} {\caii} \\
 500.1479 & & 7.505 & $-0.505$ &19 & 1.350 & 1 & 6.248 & 6.305 & 6.313 & 6.240 & 6.330 & $-0.01$ & 6.238 \\
 645.6875 & \footnote{Triplet computed as a single line; also contains two other components ($\log gf=+0.157$ and $-1.387$)}
            & 8.438 & $+0.044$ &20 & 1.850 & 1 & 6.332 & 6.383 & 6.375 & 6.297 & 6.402 & $-0.01$ & 6.322 \\
 732.3890 & \footnote{Forbidden line}& 0.000 & $-7.536$ &21 & 1.000 & 1 & 6.365 & 6.368 & 6.396 & 6.351 & 6.393 & $ 0.00$ & 6.365 \\
 824.8796 & & 7.515 & $+0.556$ &19 & 6.700 & 2 & 6.404 & 6.409 & 6.377 & 6.313 & 6.418 & $-0.07$ & 6.334 \\
 825.4721 & & 7.515 & $-0.398$ &19 & 1.800 & 2 & 6.339 & 6.372 & 6.360 & 6.301 & 6.390 & $-0.03$ & 6.309 \\
   & & & & & & & & & & & & & \\

\end{mpsupertabular}
\end{center}
\vspace{5mm}
\textbf{References:}\nopagebreak\\
\begin{minipage}[t]{0.43\linewidth}
\vspace{0pt}
\begin{enumerate}
\item Froese-Fisher \& Tachiev (\cite{FF})
\item Butler et al.\ (\cite{butl})
\item Chang \& Tang (\cite{Chang90})
\item mean of Siegel et al.\ (\cite{Siegel98}) and 2 different calculations by Froese-Fisher \& Tachiev (\cite{FF})
\item Kurucz (\cite{kur})
\item Mendoza et al.\ (\cite{mend})
\item relative values of Garz (\cite{garz}), normalised to lifetimes of O'Brian \& Lawler (\cite{obr1}, \cite{obr2})
\item mean of Schulz-Gulde (\cite{schu}), Blanco et al.\ (\cite{blan}) and Matheron et al.\ (\cite{math})
\item Berzinsh et al.\ (\cite{ber})
\item mean of Froese-Fischer \& Tachiev (\cite{FF}), Zatsarinny \& Bartschat (\cite{zats}) and Deb \& Hibbert (\cite{deb})
\item mean of Froese-Fischer \& Tachiev (\cite{FF}) and Zatsarinny \& Bartschat (\cite{zats})
\item Zerne et al.\ (\cite{zern}; values given in the body text, not
\end{enumerate}
\end{minipage}%
\begin{minipage}[t]{0.05\linewidth}
\vspace{0pt}
\hspace{3mm}
\end{minipage}%
\begin{minipage}[t]{0.43\linewidth}
\vspace{0pt}
\begin{enumerate}
\setcounter{enumi}{12}
\item[] the abstract)
\item Shabanova \& Khlyustalov (\cite{Shabanova85}), renormalised with the lifetimes of Volz \& Schmoranzer (\cite{Volz96}) and Wang et al.\ (\cite{Wang97})
\item Opacity Project calculations and assumption of $LS$ coupling (Keith Butler, via Zhang et al \cite{zhan})
\item Gamalii (\cite{Gamalii97})
\item Smith \& Raggett (\cite{smit}), as given with an extra significant figure by Smith (\cite{Smith81})
\item Smith \& Raggett (\cite{smit})
\item Smith (\cite{Smith88})
\item Opacity Project calculations and assumption of $LS$ coupling, sourced from TOPbase by Mashonkina et al (\cite{mash1})
\item as per 19, but with $LS$-coupling fine structure splittings calculated in this paper
\item Mel\'endez et al.\ (IRON Project Paper 64, \cite{melen})
\end{enumerate}
\end{minipage}
\vspace{6mm}

\begin{center}
\topcaption{\label{table:hfs} HFS data for the lines retained in this analysis}
\tablefirsthead{%
  \hline
   & \multicolumn{4}{c}{Lower level} && \multicolumn{4}{c}{Upper level} \\
  \cline{2-5}
  \cline{7-10}
  $\lambda$ & $J$ & $A$   & $B$   & HFS  && $J$ & $A$   & $B$   & HFS \\
  (nm)      &     & (MHz) & (MHz) & ref. &&     & (MHz) & (MHz) & ref. \\
  \hline
   & & & & & & & & & \\
  }
\tablehead{%
  \hline 
   & \multicolumn{4}{c}{Lower level} && \multicolumn{4}{c}{Upper level} \\
  $\lambda$ & $J$ & $A$   & $B$   & HFS  && $J$ & $A$   & $B$   & HFS \\
  (nm)      &     & (MHz) & (MHz) & ref. &&     & (MHz) & (MHz) & ref. \\
  \hline
  \multicolumn{10}{r}{continued.}\\
  \hline
   & & & & & & & & & \\
  }
\tabletail{%
  \hline
  \multicolumn{10}{r}{continued on next page}\\
  \hline
}
\tablelasttail{\hline}
\begin{mpsupertabular}{r@{\hspace{10mm}}c r r ccc r r c}
\multicolumn{10}{c} {\nai: 100\% $^{23}$Na ($I=\frac32$)} \vspace{2mm}\\
  475.1822 & $3/2$ &   18.530 & 2.721 & 1 && $1/2$ &          &       &   \\
  514.8838 & $1/2$ &   94.349 & 0.000 & 1 && $1/2$ &   37.510 & 0.000 & 2 \\ 
  615.4225 & $1/2$ &   94.349 & 0.000 & 1 && $1/2$ &   77.200 & 0.000 & 3 \\
  616.0747 & $3/2$ &   18.530 & 2.721 & 1 && $1/2$ &   77.200 & 0.000 & 3 \\
 1074.6.44 & $1/2$ &  204.300 & 0.000 & 2 && $3/2$ &    2.660 & 0.000 & 2 \\
   & & & & & \hspace{6mm} & & & & \\

\multicolumn{10}{c} {\ali: 100\% $^{27}$Al ($I=\frac52$)} \vspace{2mm}\\
  669.6023 & $1/2$ &  431.840 & 0.000 & 4 && $3/2$ &          &       &   \\  
  669.8673 & $1/2$ &  431.840 & 0.000 & 4 && $1/2$ &   20.200 & 0.000 & 5 \\
  783.5309 & $3/2$ &$-$99.000 & 0.000 & 6 && $5/2$ &          &       &   \\  
  891.2900 & $1/2$ &   58.280 & 0.000 & 7 && $3/2$ &          &       &   \\  
 1076.8365 & $1/2$ &   58.280 & 0.000 & 7 && $3/2$ &$-$72.000 & 0.000 & 8 \\
 1087.2973 & $1/2$ &   58.280 & 0.000 & 7 && $1/2$ &          &       &   \\
 1089.1736 & $3/2$ &   23.120 & 0.000 & 7 && $1/2$ &          &       &   \\
   & & & & & \hspace{6mm} & & & & \\
     	           	     	        	      	       	     		 
\multicolumn{10}{c} {\ki: 93.3\% $^{39}$K ($I=\frac32$)} \vspace{2mm}\\
 404.4142 & $1/2$ &   230.860 & 0.000 & 9 && $3/2$ &    1.973 & 0.870 &10 \\
 580.1749 & $3/2$ &     6.093 & 2.786 &11 && $1/2$ &   10.780 & 0.000 &12 \\
 693.8763 & $3/2$ &     6.093 & 2.786 &11 && $1/2$ &   21.810 & 0.000 &13 \\
1176.9639 & $3/2$ &     6.093 & 2.786 &11 && $3/2$ &    0.960 & 0.370 &14 \\
1252.2134 & $3/2$ &     6.093 & 2.786 &11 && $1/2$ &   55.500 & 0.000 &13 \\
   & & & & & \hspace{6mm} & & & & \\

\end{mpsupertabular}
\end{center}
\vspace{5mm}
\textbf{References:}\nopagebreak\\
\begin{minipage}[t]{0.43\linewidth}
\vspace{0pt}
\begin{enumerate}
\item Das \& Natarajan (\cite{Das08})
\item Safronova et al.\ (\cite{Safronova99})
\item Marcassa et al.\ (\cite{Marcassa98})
\item Nakai et al.\ (\cite{Nakai07})
\item Belfrage et al.\ (\cite{Belfrage84})
\item Otto \& Zimmermann (\cite{Otto69}), confirmed exactly by Zhao et al.\ (\cite{Zhao86})
\item Sur et al.\ (\cite{Sur05})
\end{enumerate}
\end{minipage}%
\begin{minipage}[t]{0.05\linewidth}
\hspace{3mm}
\end{minipage}%
\begin{minipage}[t]{0.43\linewidth}
\begin{enumerate}
\setcounter{enumi}{7}
\item St\"uck \& Zimmermann (\cite{Stueck70})
\item Bloom \& Carr (\cite{Bloom60}) and Dahmen \& Penselin (\cite{Dahmen67})
\item Svanberg (\cite{Svanberg71})
\item Falke et al.\ (\cite{Falke06})
\item Belin et al.\ (\cite{Belin75})
\item Gupta et al.\ (\cite{Gupta73})
\item Sieradzan et al.\ (\cite{Sieradzan97})
\end{enumerate}
\end{minipage}

\begin{table}
\caption{Adopted ionisation energies $\chi_{\rm ion}$ and partition functions $U(T)$ 
for the relevant ionisation stages of the intermediate-mass elements.}
\label{table:partition} 
\centering
\begin{tabular}{l r r r r r}
\hline
\hline Species & $E_{\rm ion}$  &  \multicolumn{4}{c}{$U(T)$} \\
               &        (eV)            & 3000\,K & 5000\,K & 8000\,K & 12000\,K \\
\hline
  \nai   &   5.139 &      2.00 &      2.00 &      3.64 &     14.08 \\
  \naii  &  47.290 &      1.00 &      1.00 &      1.00 &      1.00 \\
  \mgi   &   7.646 &      1.00 &      1.00 &      1.23 &      2.56 \\
  \mgii  &  15.040 &      2.00 &      2.00 &      2.01 &      2.09 \\
  \ali   &   5.986 &      5.80 &      5.83 &      6.20 &      8.82 \\
  \alii  &  18.830 &      1.00 &      1.00 &      1.01 &      1.10 \\
  \sii   &   8.152 &      8.61 &      9.48 &     10.35 &     13.07 \\
  \siii  &  16.350 &      5.48 &      5.68 &      5.79 &      5.95 \\
  \phosi &  10.490 &      4.04 &      4.43 &      5.46 &      7.42 \\
  \phosii&  19.770 &      7.84 &      8.63 &      9.55 &     10.52 \\
  \suli  &  10.360 &      8.29 &      8.88 &      9.61 &     10.65 \\
  \sulii &  23.340 &      4.00 &      4.12 &      4.74 &      5.96 \\
  \ki    &   4.340 &      2.00 &      2.11 &      4.66 &     17.05 \\
  \kii   &  31.630 &      1.00 &      1.00 &      1.00 &      1.00 \\
  \cai   &   6.113 &      1.00 &      1.11 &      2.57 &     11.22 \\
  \caii  &  11.870 &      2.00 &      2.19 &      2.89 &      4.25 \\
\hline
\end{tabular}
\end{table}

\appendix

\section{Hyperfine splitting}
\label{s:isohfs}

Hyperfine structure (HFS) splits the normal atomic fine-structure energy levels into sub-levels labelled by the new quantum
number $F$, which arises from the interaction of the nuclear spin $I$ and the total electron angular momentum $J$.
The effective coupling is typically of the same Russell-Saunders type as occurs between $L$ and $S$ to produce $J$.
Thus, $F$ for any given fine-structure level runs through $|J-I|\dots J+I-1, J+I$, and transitions between hyperfine
levels are permitted if $|\Delta F|\in\{0,1\}$, so long as $F=0\nleftrightarrow F'=0$.  The energies of individual HFS
levels are given (e.g.\ Pickering \cite{Pickering96a}) by
\begin{equation}
\label{hfsenergies}
E(IJF) = E(J) + A\frac{K}{2} + B\frac{3K(K+1)-4I(I+1)J(J+1)}{8I(2I-1)J(2J-1)},
\end{equation}
where
\begin{equation}
\label{K}
K = F(F+1)-J(J+1)-I(I+1).
\end{equation}
The first term in Eq.~\ref{hfsenergies} is the magnetic dipole interaction between electron and nucleus, and the second
term is the electric quadrupole interaction.  Higher multipoles can be defined, but contribute little.  $A$ and $B$ are
the HFS constants describing the respective strengths of the interactions for any given fine structure level.  We thus
determined wavelengths of hyperfine components from the selection rules and the energy shifts in Eq.~\ref{hfsenergies},
with $A$ and $B$ taken directly from experimental literature (Sect.~\ref{s:atomicdata}, where we set $B$ to zero if only $A$ was available for some level, and treated the level as unsplit if neither $A$ nor $B$ was available).  The coupling is generally strong enough that the relative intensities of the components
can also be computed using the coupling scheme (e.g.\ Morton \cite{Morton03}).  We determined $gf$-values of individual
components by scaling a line's total $gf$ via
\begin{equation}
\label{hfsgfs}
gf(IJF,IJ'F') = \frac{(2F+1)(2F'+1)}{2I+1}\bigg\{\begin{array}{ccc}J&I&F\\F'&1&J'\\\end{array}\bigg\}^2gf(J,J'),
\end{equation}
where the term in braces is the Wigner-6$j$ symbol, which we evaluated using the FORTRAN code of
Stone \& Wood (\cite{Stone80})\footnote{\href{http://www-stone.ch.cam.ac.uk/documentation/rrf/index.html}
{http://www-stone.ch.cam.ac.uk/documentation/rrf/index.html}}.


\begin{thebibliography}{}

 \bibitem[2004]{abia} Abia, C., \& Mashonkina, L. 2004, MNRAS, 350, 1127
 
 \bibitem[2001a]{alle} Allende Prieto, C., Barklem, P.~S., Asplund, M., \& Ruiz Cobo, B. 2001a, ApJ, 558, 830

 \bibitem[2001b]{cap_o} Allende Prieto, C., Lambert, D.~L., \& Asplund, M. 2001b, ApJ, 556, L63

 \bibitem[1989]{ag89} Anders, E., \& Grevesse, N. 1989, Geochim. Cosmochim. Acta, 53, 197
  
 \bibitem[2010]{andrievsky_mg} Andrievsky, S.M., Spite, M., Korotin, S.A., et al.\ 2010, A\&A, 509, A88

 \bibitem[1995]{anst} Anstee, S.~D., \& O'Mara, B.~J. 1995, MNRAS, 276, 859

 \bibitem[2000]{asp3} Asplund, M. 2000, A\&A, 359, 755 

 \bibitem[2005]{asp7} Asplund, M. 2005, ARA\&A, 43, 481

 \bibitem[1997]{asp0} Asplund, M., Gustafsson, B., Kiselman, D., \& Eriksson, K. 1997, A\&A, 318, 521

 \bibitem[2000a]{Asplund2000} Asplund, M., Ludwig, H.-G., Nordlund, \AA., \& Stein, R.~F. 2000a, A\&A, 359, 669

 \bibitem[2000b]{asp1} Asplund, M., Nordlund, \AA., Trampedach, R., Allende Prieto, C., \& Stein, R. F. 2000b, A\&A, 359, 729

 \bibitem[2004]{asp4} Asplund, M., Grevesse, N., Sauval, A.~J., Allende Prieto, C., \& Kiselman, D., 2004, A\&A, 417, 751

 \bibitem[2005]{AGS05} Asplund, M., Grevesse, N., \& Sauval, A.~J. 2005, in ASP Conf.\ Ser.\ 336, ed. T. G. Barnes III \& F. N. Bash (Astron.\ Soc.\ Pac., San Francisco), 25

 \bibitem[2009]{asp8} Asplund, M., Grevesse, N., Sauval, A.~J., \& Scott, P. 2009, ARA\&A, 47, 481 (AGSS09)

 \bibitem[2007]{bark2} Barklem, P.~S. 2007, AIPC, 938, 111

 \bibitem[2000]{bark1} Barklem, P.~S., Piskunov, N., \& O'Mara, B. J. 2000, A\&AS, 142, 467

 \bibitem[2003]{barklem03} Barklem P.~S., Belyaev A. K., \& Asplund M. 2003, A\&A, 409, L1

 \bibitem[2010]{barklem_na} Barklem, P.~S., Belyaev, A.~K., Dickinson, A.~S., \& Gadea, F.~X., 2010, A\&A, 519, A20

 \bibitem[2012]{barklem_mg} Barklem, P.~S., Belyaev, A.K., Spielfiedel, A., Guitou, M., \& Feautrier, N., 2012, A\&A, 541, A80

 \bibitem[1998]{bau1} Baum\"uller, D., Butler, D., \& Gehren, T. 1998, A\&A, 338, 637

 \bibitem[1996]{bau2} Baum\"uller, D., \& Gehren, T. 1996, A\&A, 307, 961
 
 \bibitem[2012]{beeck2012} Beeck, B., Collet, R., Steffen, M., et al.\ 2012, A\&A, 539, 121

 \bibitem[1984]{Belfrage84} Belfrage, C., H\"orb\"ack, S., Levinson, C., et al.\ 1984, Z. Phys. A, 316, 15

 \bibitem[1975]{Belin75} Belin, G., Holmgren, L., Lindgren, I., \& Svanberg, S. 1975, Phys. Scr, 12, 287

 \bibitem[2013]{belyaev_al} Belyaev, A.~K., 2013, A\&A, 560, A60
 
 \bibitem[2012]{Bergemann2012} Bergemann, M., Lind, K., Collet, R., Magic, Z. \& Asplund, M. 2012, MNRAS, 427, 27

 \bibitem[1997]{ber} Berzinsh, U., Svanberg, S., \& Bi\'emont, E. 1997, A\&A, 326, 412

 \bibitem[1995]{blan} Blanco, F., Botho, B., \& Campos, J. 1995, Phys. Scr., 52, 628

 \bibitem[1960]{Bloom60} Bloom, A. L., \& Carr, J. B. 1960, Phys. Rev., 119, 1946

 \bibitem[1993]{butl} Butler, K., Mendoza, C., \& Zeippen, C. J. 1993, J. Phys. B, 26, 4409

 \bibitem[2007]{caf3} Caffau, E., \& Ludwig, H.-G. 2007, A\&A, 467, L11

 \bibitem[2007a]{caf2} Caffau, E., Faraggiana, R., Bonifacio, P., Ludwig, H.-G., \& Steffen, M. 2007a, A\&A, 470, 699

 \bibitem[2007b]{caf1} Caffau, E., Steffen, M., Sbordone, L., Ludwig, H.-G., \& Bonifacio, P.  2007b, A\&A, 473, L9

 \bibitem[2011]{caf4} Caffau, E., Ludwig, H.-G., Steffen, M., Freytag, B., \& Bonifacio, P.  2011, Sol. Phys., 268, 255

 \bibitem[1986]{multi} Carlsson, M. 1986, Uppsala Astron. Obs. Rep. 33

 \bibitem[1990]{Chang90} Chang, T. N. \& Tang, X.  1990, JQSRT, 43, 207

 \bibitem[1967]{Dahmen67} Dahmen, H. L., \& Penselin, S. 1967, Z. Phys., 200, 456

 \bibitem[2008]{Das08} Das, D., \& Natarajan, V.  2008, J. Phys. B, 41, 035001

 \bibitem[2008]{deb} Deb, N. C., \& Hibbert, A. 2008, At. Dat. Nucl. Dat. Tab., 94, 561

 \bibitem[1973]{delb1} Delbouille, L., Neven, L., \& Roland, G. 1973, Atlas photom{\'e}trique du spectre solaire de $\lambda 3000$ a $\lambda 10000$ (Institut d'Astrophysique, Universit\'e de Li\`ege)

 \bibitem[1981]{delb2} Delbouille, L., Roland, G., Brault, J. W., \& Testerman, L. 1981, Photometric Atlas of the Solar Spectrum from 1850 to 10000\,cm$^{-1}$ (Kitt Peak National Observatory, Tucson)
 
 \bibitem[1990]{stell-gran4} Dravins, D. \& Nordlund, {\AA}. 1990, A\&A, 228, 184

 \bibitem[1968]{dra} Drawin, H. W. 1968, Z. Phys., 211, 404

 \bibitem[2006]{Falke06} Falke, S., Tiemann, E., Lisdat, C., Schnatz, H., \& Grosche, G. 2006, Phys. Rev. A, 74, 032503

 \bibitem[2011]{FF} Froese-Fisher, C. \& Tachiev, G. 2011 (\href{http://nlte.nist.gov/MCHF/}{http://nlte.nist.gov/MCHF/})

 \bibitem[1997]{Fuhrmann97} Fuhrmann, K., Pfeiffer, M., Frank, C., Reetz, J., \& Gehren, T. 1997, A\&A, 323, 909

 \bibitem[1997]{Gamalii97} Gamalii, V. F. 1997, Opt. Spectros., 83, 662

 \bibitem[1973]{garz} Garz, T. 1973, A\&A, 26, 471

 \bibitem[2004]{gehr} Gehren, T., Liang, Y. C., Shi, J. R., Zhang, H. W., \& Zhao, G. 2004, A\&A, 413, 1045
 
 \bibitem[1960]{goldberg60} Goldberg, L., M\"uller, E.A., \& Aller, L.H. 1960, ApJS, 5, 1

 \bibitem[1998]{gs98} Grevesse, N., \& Sauval, A.J. 1998, Space Sci. Rev., 85, 161
 
 \bibitem[2014]{AGSS_heavy} {Grevesse}, N., {Scott}, P., {Asplund}, M., \& Sauval, A.~J. 2014, A\&A in press, arXiv:1405.0288 (Paper III)

 \bibitem[1973]{Gupta73} Gupta, R., Happer, W., Lam, L. K., \& Svanberg, S. 1973, Phys. Rev. A, 8, 2792

 \bibitem[1975]{bgus:atmgrid} Gustafsson, B., Bell, R. A., Eriksson, K. \& Nordlund, {\AA}. 1975, A\&A, 42, 407

 \bibitem[2008]{marcs08} Gustafsson, B., Edvardsson, B., Eriksson, K., et al.\ 2008, A\&A, 486, 951

 \bibitem[1969]{hall_f} Hall, D.N.B., \& Noyes, R.W. 1969, Ap. Letters, 4, 143

 \bibitem[1972]{hall_cl} Hall, D.N.B., \& Noyes, R.W. 1972, ApJ, 175, L95

 \bibitem[1967]{hol0} Holweger, H. 1967, Z. f. Astrophys., 65, 365

 \bibitem[1979]{hol1} Holweger, H. 1979, in The elements and their isotopes in the Universe, 22nd Li\`ege International Astrophysical Colloquium, Institut d'Astrophysique, University of Li\`ege, Belgium, 117

 \bibitem[1974]{holm} Holweger, H., \& M\"{u}ller, E. A. 1974, Sol. Phys., 39, 19

 \bibitem[2014]{Jonsson14} J{\"o}nsson, H., Ryde, N., {Harper}, G.~M., et al.\ 2014, A\&A, 564, A122

 \bibitem[2008a]{kel1} Kelleher, D. E., \& Podobedova, L. I. 2008a, J. Phys. Chem. Ref. Data, 37, 267

 \bibitem[2008b]{kel2} Kelleher, D. E., \& Podobedova, L. I. 2008b, J. Phys. Chem. Ref. Data, 37, 709

 \bibitem[1993]{atlas9} Kurucz, R. L. 1993, Kurucz CD-ROM 13 (Harvard-Smithsonian Center for Astrophysics) 

 \bibitem[2011]{kur} Kurucz, R. L. 2011 (\href{http://kurucz.harvard.edu/}{http://kurucz.harvard.edu/})

 \bibitem[1978]{lamb1} Lambert, D. L., \& Luck, R. E., 1978, MNRAS, 183, 79
 
 \bibitem[2011]{lind_na} Lind, K., Asplund, M., Barklem, P.S., \& Belyaev, A.K., 2011, A\&A, 528, A103

 \bibitem[2013]{lind_li6} Lind, K., Mel\'endez, J., Asplund, M., Collet, R., \& Magic, Z., 2013, A\&A, 554, A96

 \bibitem[2003]{lodd03} Lodders, K. 2003, ApJ, 591, 1220

 \bibitem[2009]{lodd} Lodders, K., Palme, H., \& Gail, H.-P. 2009, Landolt B\"ornstein, New Series, Vol.\ VI/4B, Chap.\ 4.4, Abundances of the Elements in the Solar System, ed. J.~E.\ Tr\"umper (Springer-Verlag, Berlin), 560--630

 \bibitem[2013]{magic_stagger} {Magic}, Z., {Collet}, R., {Asplund}, M., et al.\ 2013, A\&A, 557, A26

 \bibitem[2014]{maiorca} {Maiorca}, E., {Uitenbroek}, H., {Uttenthaler}, S., et al.\ 2014, ApJ, 788, 149

 \bibitem[1998]{Marcassa98} Marcassa, L. G., Muniz, S. R., Telles, G. D., Zilio, S. C., Bagnato, V. S. 1998, Opt. Comm., 155, 38

 \bibitem[2013]{mashonkina_mg} Mashonkina, L.I., 2013, A\&A, 550, A28

 \bibitem[2007]{mash1} Mashonkina, L. I., Korn, A. J., \& Przybilla, N. 2007, A\&A, 461, 261

 \bibitem[2001]{math} Matheron, P., Escarguel, A., Redon, R., Lesage, A., \& Richou, J. 2001, JQSRT, 69, 535

 \bibitem[2007]{melen} Mel\'endez, M., Bautista, M. A., \& Badnell, N. R. 2007, A\&A, 469, 1203

 \bibitem[1995]{mend} Mendoza, C., Eissner, W., LeDourneuf, M., \& Zeippen, C. J. 1995, J, Phys. B, 28, 3485

 \bibitem[2003]{Morton03} Morton, D. C. 2003, ApJS, 149, 205

 \bibitem[2007]{Nakai07} Nakai, H., Jin, W.-G., Kawamura, et al.\ 2007, Jap. J. App. Phys., 46, 815

 \bibitem[1984]{neck} Neckel, H., \& Labs, D. 1984, Sol. Phys., 90, 205
 
 \bibitem[1982]{nordlund82} Nordlund, \AA. 1982, A\&A, 107, 1

 \bibitem[1995]{stagger} Nordlund, \AA., \& Galsgaard, K. 1995, A 3D MHD Code for Parallel Computers, Technical Report, Astronomical Observatory, Copenhagen University

 \bibitem[2009]{nord} Nordlund, \AA., Stein, R., \& Asplund, M. 2009, Living Reviews in Solar Physics, 6, 2
 
 \bibitem[1991a]{obr1} O'Brian, T. R. \& Lawler, J. E. 1991a, Phys. Lett. A, 152, 407

 \bibitem[1991b]{obr2} O'Brian, T. R. \& Lawler, J. E. 1991b, Phys. Rev. A, 44, 7134

 \bibitem[1969]{Otto69} Otto, H. \& Zimmermann, P. 1969, Z. Phys., 225, 269

 \bibitem[2009]{pereira_o} Pereira, T. M. D., Asplund, M., \& Kiselman, D. 2009, A\&A, 508, 1403

 \bibitem[2013]{pereira_models} Pereira, T. M. D., Asplund, M., Collet, R., et al.\ 2013, A\&A, 554, A118

 \bibitem[{1996}]{Pickering96a} {Pickering}, J.~C. 1996, \apjs, 107, 811

 \bibitem[1998]{IUPAC98} Rosman, K. J. R. \& Taylor, P. D. P.  1998, Pure \& Appl. Chem., 70, 217
 
 \bibitem[1929]{russell29} Russell, H.N. 1929, ApJ, 70, 11

 \bibitem[1999]{Safronova99} Safronova, M. S., Johnson, W. R., \& Derevianko, A.  1999, Phys. Rev. A, 60, 4476

 \bibitem[2008]{Sansonetti08} Sansonetti, J. E. 2008, J. Phys. Chem. Ref. Data 37, 7

 \bibitem[1969]{schu} Schulz-Gulde, E. 1969, JQSRT, 9, 13

 \bibitem[2014]{AGSS_Fegroup} Scott, P., Asplund, M., Grevesse, N., Sauval, A.~J., \& Bergemann, M. 2014, A\&A in press, arXiv:1405.0287 (Paper II)

 \bibitem[1985]{Shabanova85} Shabanova, L. N. \& Khlyustalov, A. N. 1985, Opt. Spectros., 59, 123

 \bibitem[2004]{shi1} Shi, J. R., Gehren, T., \& Zhao, G. 2004, A\&A, 423, 683

 \bibitem[2008]{shi2} Shi, J. R., Gehren, T., Butler, K., Mashonkina, L., \& Zhao, G. 2008, A\&A, 486, 303

 \bibitem[1998]{Siegel98} Siegel, W., Migdalek, J. \& Kim, Y.-K.  1998, At. Data Nuc. Data Tables, 68, 303

 \bibitem[1997]{Sieradzan97} Sieradzan, A., Stoleru, R., Yei, W., \& Havey, M. D. 1997, Phys. Rev. A, 55, 3475

 \bibitem[1981]{Smith81} Smith, G. 1981, A\&A, 103, 351

 \bibitem[1988]{Smith88} Smith, G. 1988, J. Phys. B, 21, 2827

 \bibitem[1981]{smit} Smith, G. \& Raggett, D. J. 1981, J. Phys. B, 14, 4015
 
  \bibitem[1980]{Stone80} {Stone}, A.~J. \& {Wood}, C.~P. 1980, Comput. Phys. Commun., 21, 195

 \bibitem[1970]{Stueck70} St\"{u}ck, H. L., \& Zimmermann, P. 1970, Z. Phys., 239, 345

 \bibitem[1956]{suess56} Suess, H.E., \& Urey, H.C. 1956, Rev. Mod. Phys, 28, 53

 \bibitem[2005]{Sur05} Sur, C., Chaudhuri, R. K., Das, B. P., \& Mukherjee, D. 2005, J. Phys. B, 38, 4185

 \bibitem[1971]{Svanberg71} Svanberg, S. 1971, Phys. Scr, 4, 275

 \bibitem[1996]{tak2} Takeda, Y., Kato, K., Watanabe, Y, \& Sadakane, K. 1996, PASJ, 48, 511

 \bibitem[2005]{tak1} Takeda, Y., Hashimoto, O., Taguchi, H., et al.\ 2005, PASJ, 57, 751

 \bibitem[2013]{trampedach:3Datmgrid} Trampedach, R., Asplund, M., Collet, R., Nordlund, {\AA}, \& Stein, R.~F. 2013, ApJ, 769, 18

 \bibitem[1955]{unsold} Uns\"old, A. 1955, {Physik der Sternatmospharen, mit besonderer Berucksichtigung der Sonne.}, 2nd edn. (Springer, Berlin)

 \bibitem[1996]{Volz96} Volz, U. \& Schmoranzer, H. 1996, Phys. Scr, T65, 48

 \bibitem[1997]{Wang97} Wang, H., Li, J., Wang, X. T., Williams, C. J., Gould, P. L., Stwalley, W. C. 1997, Phys. Rev. A, 55, R1569

 \bibitem[2001]{wede} Wedemeyer, S. 2001, A\&A, 373, 998

 \bibitem[2006]{zats} Zatsarinny, O., \& Bartschat, K. 2006, J. Phys. B, 39, 2861

 \bibitem[1997]{zern} Zerne, R., Caiyan, L., Berzinsh, U., \& Svanberg, S. 1997, Phys. Scr, 56, 459

 \bibitem[2006]{zhan} Zhang, H. W., Butler, K., Gehren, T., Shi, J. R., \& Zhao, G. 2006, A\&A, 453, 723

 \bibitem[1986]{Zhao86} Zhao, Y. Y., Carlsson, J., Lundberg, H., \& Wahlstr\"om, C. G. 1986, Z. Phys. D, 3, 365

 \bibitem[1998]{zhao1} Zhao, G., Butler, K., \& Gehren, T. 1998, A\&A, 333, 219

\end{thebibliography}
\end{document}